\documentclass[a4paper,12pt]{article}
\usepackage{amsmath,amssymb}
\usepackage{hyperref}
\usepackage{graphicx}
\usepackage{mathrsfs}
\usepackage{float}
\usepackage{ulem}
\hypersetup{
    colorlinks,%
    citecolor=blue,%
    filecolor=blue,%
    linkcolor=blue,%
   urlcolor=blue,
   linktoc=page
}
\usepackage{chngcntr}
\counterwithin{equation}{section}
\usepackage{cite}

\newcommand{\be}{\begin{equation}}
\newcommand{\ee}{\end{equation}}
\newcommand{\beq}{\begin{equation}}
\newcommand{\eeq}{\end{equation}}
\newcommand{\f}{\frac}
\newcommand{\s}{\sqrt}
\newcommand{\p}{\partial}

\newcommand{\om}{\mathcal{O}}
\newcommand{\bea}{\begin{equation}\begin{aligned}}
\newcommand{\eea}{\end{aligned}\end{equation}}
\newcommand{\ba}{\begin{align}}
\newcommand{\ea}{\end{align}}

\newcommand{\ra}{\rangle}
\newcommand{\la}{\langle}
\begin{document}

\begin{titlepage}

\vspace{.4cm}
\begin{center}
\noindent{\Large \textbf{Modular Hamiltonians of excited states, OPE blocks and emergent bulk fields}}\\
\vspace{1cm}
G\'abor S\'arosi$^{a,b,}$\footnote{gsarosi@vub.ac.be}
and
 Tomonori Ugajin$^{c,}$\footnote{ugajin@kitp.ucsb.edu}

\vspace{.5cm}
 {\it
 $^{a}$Theoretische Natuurkunde, Vrije Universiteit Brussels and \\ International Solvay Institutes,\\
Pleinlaan 2, Brussels, B-1050, Belgium\\
\vspace{0.2cm}
 }
 \vspace{.5cm}
  {\it
  $^{b}$David Rittenhouse Laboratory, University of Pennsylvania,\\
  Philadelphia, PA 19104, USA\\
\vspace{0.2cm}
 }
\vspace{.5cm}
  {\it
 $^{c}$Kavli Institute for Theoretical Physics, University of California, \\
Santa Barbara, 
CA 93106, USA\\
\vspace{0.2cm}
 }
\end{center}


\begin{abstract}
We study the entanglement entropy and the modular Hamiltonian of slightly excited states reduced to a ball shaped region in generic conformal field theories. We set up a formal expansion in the one point functions of the state in which all orders are explicitly given in terms of integrals of multi-point functions along the vacuum modular flow, without a need for replica index analytic continuation. We show that the quadratic order contributions in this expansion can be calculated in a way expected from holography, namely via the bulk canonical energy for the entanglement entropy, and its variation for the modular Hamiltonian. The bulk fields contributing to the canonical energy are defined via the HKLL procedure. In terms of CFT variables, the contribution of each such bulk field to the modular Hamiltonian is given by the OPE block corresponding to the dual operator integrated along the vacuum modular flow. These results do not rely on assuming large $N$ or other special properties of the CFT and therefore they are purely kinematic. 
\end{abstract}

\end{titlepage}

\tableofcontents
\pagebreak

\section{Introduction}

The principal tool to quantify the amount of entanglement in a pure state over some bipartition is the entanglement entropy, the von Neumann entropy of the reduced density matrix \cite{nielsen2010quantum}. It is fair to say that this information theoretic quantity has took a central stage in contemporary theoretical high energy and condensed matter physics. A far from exhaustive list includes its fundamental role in recent studies of renormalization group flows \cite{Casini:2004bw,Casini:2012ei,Casini:2016fgb,Casini:2016udt,Casini:2017vbe}, quantum quenches \cite{Calabrese:2007mtj,Calabrese:2016xau,Caputa:2016yzn,Caputa:2014vaa,Hartman:2013qma,Takayanagi:2010wp,Mandal:2016cdw,Ugajin:2013xxa}, and quantum gravity \cite{Ryu:2006bv,Ryu:2006ef,VanRaamsdonk:2010pw,Lewkowycz:2013nqa,Faulkner:2013ana}. However, explicit calculations of the entanglement entropy in a field theoretic setup are available mostly only for the vacuum state of conformal field theories \cite{Holzhey:1994we,Calabrese:2004eu,Calabrese:2009qy,Calabrese:2009ez,Calabrese:2010he,Casini:2011kv,Wong:2013gua}.

A more refined object, containing information about both entanglement and other statistical properties, such as distinguishability, is the so called modular Hamiltonian (or entanglement Hamiltonian), which is minus the logarithm of the reduced density matrix. This quantity has proved to be very useful in various field theoretic \cite{Casini:2008cr,Faulkner:2016mzt} and gravitational \cite{Wall:2011hj,Faulkner:2013ica,Bousso:2014sda,Bousso:2014uxa,Lashkari:2016idm,Kabat:2017mun} contexts, despite the fact that most generally in relativistic field theory its explicit form is only known for the vacuum state reduced to half space. The situation is much better for conformal field theories and their small deformations, where ball shaped regions \cite{Casini:2011kv}, deformed half spaces \cite{Faulkner:2016mzt} and regions with boundaries in a null plane (or cone) \cite{Casini:2017roe} are accessible. There are also a number of additional situations that are tractable for the vacuum in two dimensions \cite{Cardy:2016fqc}. Much less is explicitly known about other states than the vacuum, where the modular Hamiltonian is a heavily nonlocal operator, see however \cite{Jafferis:2014lza,Klich:2015ina,Lashkari:2015dia,Klich:2017qmt} for some examples.

\vspace{4mm}

In this paper, we present a novel way of calculating the entanglement entropy and the modular Hamiltonian of excited states, reduced to a single ball shaped region, in conformal field theories living on the cylinder $\mathbb{R} \times S^{d-1}$. 
In order to do this, we first develop  a general prescription to calculate the modular Hamiltonian $K_{\rho}$ of a density matrix $\rho$ which has the form $\rho= \rho_{0}+\delta \rho$, where $\rho_{0}$ is a reference density matrix whose modular Hamiltonian $K_0=-\log \rho_0$ is known.  The result is schematically written as
\be
K_{\rho} =K_0 + \sum_{n=1}^{\infty} (-1)^{n} \int_{-\infty}^\infty ds_1...ds_n \mathcal{K}_n(s_1,...,s_n)\prod_{k=1}^n (e^{-\left( \frac{is_k}{2\pi}+\frac{1}{2}\right) K_0}\delta \rho e^{\left(\frac{is_k}{2\pi}-\frac{1}{2}\right)K_0}), 
\ee
where the kernel $ \mathcal{K}_n$ is going to be given explicitly in the bulk of the paper. On the right hand side we have a product of $n$ copies of $\delta \rho$s with $n$ integrals with respect to the modular flow generated by the reference density matrix $\rho_{0}$.\footnote{Our expansion is similar in spirit to the one considered in \cite{Rosenhaus:2014woa,Kelly:2015mna}, the new point here is that we recast all orders in the form of modular integrals.} The von Neumann entropy  $S(\rho)$ is then simply computed by the expectation value of the modular Hamiltonian $K_{\rho}$.  This formula is quite general, and applicable to any theory, since we only use operator identities to derive it.

We then apply the above formula to conformal field theories, taking $\rho$ to be the reduced density matrix of an excited state.  This will result in a formal 
series expansion for the ``first law substracted"\footnote{By ``first law substracted" we mean that we substract the expectation value of the vacuum modular Hamiltonian. This way, the expansion is really for the relative entropy $S(\rho_V||\rho_{\rm vac})$.} entanglement entropy. We say that the expansion is formal, because there are some signs, which we will discuss, that for CFTs it actually does not converge.\footnote{We can formally introduce a small parameter $t$ whose powers count the orders in our expansion by considering the relative entropy $S\big((1-t)\rho_{\rm vac}+t \rho_V||\rho_{\rm vac}\big)$. We will present some evidence that the series for this has zero radius of convergence. Nevertheless, each order can be straightforwardly re-expanded in some different small parameter of interest, e.g. the radius of the subsystem, such that the reorganized series have no obstacle to be convergent.}
In a CFT, the $\delta \rho$ of interest is given by a sum of OPE blocks of primary operators. 
 From this observation we will show that the $n$th order term in the expansion applied to the entanglement entropy 
is a $2n+2$ point function of the operator $V$, creating the excited state $|V\rangle$ in question, integrated along the vacuum modular flow. There is a recent grow in interest \cite{Cotler:2017erl,Faulkner:2017vdd} in such integrals in connection with bulk reconstruction \cite{Hamilton:2006az,Czech:2012bh,Dong:2016eik} in AdS/CFT \cite{Maldacena:1997re,Witten:1998qj}, therefore we expect our result to be a very natural one from a holographic point of view. 

Indeed, we will derive several bulk quantities from these CFT results. First of all, by rewriting the contribution of the conformal family of a given primary $\om$ to the quadratic order piece of the entanglement entropy $S^{(2)}_{V}$, we obtain the bulk canonical energy \cite{Hollands:2012sf} of the field profile $\langle V|\phi |V\rangle$, in a spatial slice $\Sigma$ of the entanglement wedge in vacuum AdS. The profile $\langle V|\phi |V\rangle$ is associated to a bulk field operator $\phi$ defined via the HKLL procedure \cite{Hamilton:2006az} from $\om$. Explicitly, we will show that
\be
S^{(2)}_{V}=-2\pi \int_{\Sigma} d\Sigma^{a} \xi^{b} T_{ab} (\langle V|\phi |V\rangle),
\ee
where $\xi^{a}$ is the Killing vector generating the bulk vacuum modular flow, and $T_{ab} (\phi)$ is the stress tensor of the free bulk field $\phi$. Notice that the expectation value is inside the function so the expression is indeed quadratic in $\langle V|\phi |V\rangle$.
This quadratic order piece is what determines the so called quantum Fisher information in the vacuum. Therefore we show that the result in \cite{Lashkari:2015hha,Nozaki:2013vta,Lin:2014hva,Beach:2016ocq} formally holds for \textit{any} CFT, without assuming anything about the theory being holographic.\footnote{One can think of this statement as the Neumann boundary condition version of the result in \cite{Faulkner:2014jva} for the Dirichlet case.} In particular we did not use the Ryu Takayangi formula or its generalizations to derive this statement. 

 We will also briefly discuss the implications of this observation for the modular Hamiltonian and the relative entropy. On the CFT side, the leading contribution to the modular Hamiltonian of an excited state is given by the global OPE block $B_{\om}(x_{1}, x_{2})$ of $\om$ integrated along the vacuum modular flow with a specific kernel. This, in turn, can be rewritten as a linear variation of the canonical energy in the bulk field dual to $\om$, which is a bulk codimension one integral of the bulk field operator with some smearing function. This leads us to the following identity
\be
\label{eq:opeblocksurfaceint}
C^\om_{VV} \int^{\infty}_{-\infty} ds \f{B_{\om}(\tau-\pi+is,\hat{\tau}-\pi+is)}{(\cosh \f{s}{2})^2} =2\pi \int_{\Sigma}d\Sigma^{a} \xi^{b} \tilde{T}_{ab} (\la V|\phi | V\ra, {\phi}),
\ee
relating the integral of the OPE block along the vacuum modular flow to the integral of the dual bulk local operator ${\phi}$ on the slice $\Sigma$ of the entanglement wedge in AdS.  Here, $\tilde{T}_{ab}$ is a bulk tensor related to the variation of stress energy while $\tau$ and $\hat \tau$ are Euclidean Rindler times depending on the size of the ball. These quantities will be defined in the bulk of the paper.
The relation \eqref{eq:opeblocksurfaceint} is reminiscent to the bulk surface operator/OPE block story of \cite{Czech:2016xec,deBoer:2016pqk}. In addition, this allows us to write the leading contribution of the conformal family of $\om$ to the modular Hamiltonian as
\beq
K_V = \frac{A}{4G} + K_V^{\rm bulk} + \cdots,
\eeq
where $A$ is the perturbative area operator around empty AdS, while $K_V^{\rm bulk}$ is -- up to a constant -- the bulk modular Hamiltonian of the dual field $\phi$ in the coherent state associated to the classical profile $\langle V| \phi |V\rangle$. This is a special version of the JLMS conjecture \cite{Jafferis:2015del} which is therefore valid in any CFT.

As a simple application of this modular Hamiltonian, we also give a holographic formula for the quadratic part of the  relative entropy $S(\rho_{V} || \rho_{W})$  of slightly excited states $\rho_{V}, \rho_{W}$ 
\be
S(\rho_{V} || \rho_{W})= 2\pi \int_{\Sigma} d\Sigma^{a} \xi^{b} T_{ab} (\la\phi \ra_{V} -\la\phi \ra_{W}) + \cdots,
\ee
with $\la\phi \ra_{V} = \la V| \phi |V \ra, \la\phi \ra_{W} = \la W| \phi |W \ra$ and $\cdots$ denoting similar contributions from other bulk fields.

\vspace{4mm}

In addition to its suggestive form, our expansion should also be useful as a technical tool. In particular, when one is interested in the entanglement entropy up to some given order in the radius of the subsystem, the series can be truncated.
The advantage compared to the replica trick \cite{Calabrese:2009qy} is that there is no need to perform a problematic\footnote{These analytic continuations can be very hard to do at higher powers of the radius, not to mention that they are ambiguous due to Carlson's theorem \cite{Calabrese:2010he}.} analytic continuation in the replica index, instead, each order can be given by a well-defined multidimensional integral which in principle can be evaluated numerically. Still, the result is applicable to any interacting CFT, unlike other methods not involving an analytic continuation \cite{Casini:2009sr}. As a demonstration, we will reproduce some of the earlier results \cite{Sarosi:2016oks,Sarosi:2016atx,Ugajin:2016opf,He:2017vyf} in the small subsystem size expansion, some of which only had a holographic derivation so far.

Another situation where our expansion is natural is for single trace operators in holography. The $2n+2$ point functions appearing at order $n$ are proportional to the product of $n+1$ operator product expansion (OPE) coefficients $C^\om_{VV}$ of the operator $V$ with other local operators of the theory. When the excited state $|V\rangle$ in question is a perturbative supergravity excitation above the AdS vacuum, most of these OPE coefficients are suppressed at least as
$\frac{1}{N}$. The only exception to this are OPE coefficients into ``multitraces" of $V$, which produce the leading order generalized free field result. Therefore, the bulk perturbation theory results to any given order in $\frac{1}{N}$ can in principle be extracted from our formula, with the important caveat that actually doing this would require the resummation of the generalized free field part of the expansion. We will not attempt to make progress in this direction in the present paper, but we will comment a little more on this later on.

\vspace{4mm}

This paper is organized as follows. In section \ref{sec:modham} we recast the series expansion of the noncommutative $\-\log(\rho_0+\delta \rho)$ in a way that involves integrals of $\delta \rho$ translated along the modular flow of $\rho_0$, and apply this to the Rindler representation of the reduced density matrix of the excited state to obtain an expansion for its modular Hamiltonian. In section \ref{ref:ee} we apply these results to obtain the announced expansion of the entanglement entropy. In the remainder of this section, we somewhat simplify the first two terms of the expansion, confirm that the leading order piece gives an identical answer as the replica trick, and finally identify it with the quantum Fisher information in the vacuum. In section \ref{sec:holo} we present the holographic rewriting of the leading order piece in terms of the canonical energy, and discuss the holographic form of the modular Hamiltonian and the relative entropy at this order. Section \ref{sec:smallsub} involves some explicit calculations with the new method in the small subsystem size limit. We conclude the paper and discuss some possible future directions in section \ref{sec:discuss}.

\section{Modular Hamiltonian}
\label{sec:modham}

\subsection{The Casini-Huerta-Myers map}

Let us consider a $d$ dimensional CFT on the cylinder $\mathbb{R}\times S^{d-1}$ with (Lorentzian) metric
\be
ds^2= -dt^2 +d\theta^2 +\sin ^{2} \theta d\Omega_{d-2}^2.
\ee 
We are interested in the reduced density matrix of excited states to the cap-like region
\beq
\label{eq:regionA}
A: \big( t=0, \theta \in [0,\theta_0], \Omega_{d-2} \big).
\eeq
The casual diamond of this region can be mapped to Lorentzian $\mathbb{R} \times H^{d-1}$ by picking an analogue of Rindler coordinates, developed in \cite{Casini:2011kv}. The Euclidean version of this map with $t_E=it$ maps the entire Euclidean cylinder to $S^1 \times H^{d-1}$, with the hyperbolic space $H^{d-1}$ and the $S^1$ having equal radius. This map reads explicitly as
\be
\label{eq:CHM}
\tanh t_E =\f{\sin \theta_{0} \sin \tau}{\cosh u+ { \cos \theta_0}\cos \tau}, \qquad \tan \theta =\f{\sin \theta_{0}\sinh u}{\cos \tau+\cos \theta_{0} { \cosh u} },
\ee
whereas the metric on Euclidean $S^1 \times H^{d-1}$ is
\be
ds^2_{\Sigma_{n}}= d\tau^2 + du^2+{ \sinh^2 u d\Omega_{d-2}^2},
\ee
where $\tau$ is the Euclidean time of $S^1$ with the identification $\tau \sim \tau +2\pi$ and $u$ is the radial coordinate of hyperbolic space. We will refer to this specific map as the Casini-Huerta-Myers (CHM) map.

\subsection{Rindler representation of the density operator}

Let us consider an excited state $|V\rangle$ on the cylinder $\mathbb{R} \times S^{d-1}$. The Euclidean cylinder is conformal to flat space via $t_E=e^r$, where $r$ is the radial coordinate. Therefore, we can prepare the state by doing the Euclidean path integral in the unit ball of this flat space, with a local operator $V$ inserted at the origin. This is the standard radial quantization, and it leads to the state-operator correspondence of the CFT. Now let us consider the reduced density matrix of $|V\rangle$ to region $A$ of \eqref{eq:regionA}
\beq
\rho_V = \text{Tr}_{A^c} |V\rangle \langle V|.
\eeq
We can prepare matrix elements $\langle .|\rho_V|.\rangle$ of this reduced density matrix by mapping it with the CHM map \eqref{eq:CHM} to the Euclidean path integral on $S^1\times H^{d-1}$, with insertions of the image of two copies of the operator $V$ at the center of hyperbolic space and Euclidean time coordinates
\bea
\label{eq:globalinsertion}
\tau &=\pi-\theta_0 && \text{and} && \hat \tau &=\pi+\theta_0.
\eea
We can cut open the space along the $S^1$ and pass over to a Hamiltonian picture. This gives us the density operator in a Rindler representation
\bea
\rho_V &\sim e^{-(\pi-\theta_0)K} V(0) e^{-2\theta_0 K} V(0) e^{-(\pi-\theta_0)K} \\
&=e^{-\pi K} V(-\theta_0)V(\theta_0)e^{-\pi K},
\eea
where $K$ is the modular Hamiltonian of the vacuum, or equivalently, the translation generator in the Lorentzian $-i\tau$ direction.\footnote{We slightly break our conventions with the vacuum with the extra $2\pi$ factor in $-\log\rho_{\rm vac}=2\pi K$. This is so that $K$ is the translation generator on a circle of unit radius.} Here and bellow, we suppress dependence of local operators on the coordinates of $H^{d-1}$ for simplicity. In the present context of globally excited states, all operators sit at the center of hyperbolic space. 
We interpret this as the reduced density operator of the state $|V\rangle$ to the cap like region $A$, in the Rindler conformal frame.
To set the normalization right, i.e. $\text{Tr}\rho_V=1$, we need to divide by the thermal two point function\footnote{We assume the vacuum modular Hamiltonian to be normalized such that $\text{Tr}e^{-2\pi K}=1$.}:
\bea
\label{eq:rhoV}
\rho_V &=\frac{ e^{-\pi K} V( \tau-\pi)  V(\hat \tau-\pi)e^{-\pi K} }{\langle V(\tau)V(\hat \tau)\rangle_{S^1\times H^{d-1}}} .
\eea

When $V$ is a conformal primary, we can use the $V\times V$ OPE in \eqref{eq:rhoV} 
\beq
\label{eq:OPE}
V( \tau-\pi)  V(\hat\tau-\pi) = \langle V(\tau)V(\hat \tau)\rangle_{S^1\times H^{d-1}} \sum_k C^k_{VV} B_k( \tau-\pi, \hat \tau-\pi).
\eeq
Here, $C^k_{VV}\equiv C^{\om_k}_{VV}$ are OPE coefficients, and $B_k\equiv B_{\om_k}$ are OPE blocks on $S^1\times H^{d-1}$ , summing up all the conformal descendants of the conformal primary $\om_k$. They can be formally written as
\be 
C^k_{VV} B_{k} (x_{1}, x_{2}) \equiv  C( x_{12} , \p_{2}) \om_k (x_{2}), \label{eq:OPB}
\ee
where $C( x_{12} , \p_{2})$ is some differential operator fixed by conformal invariance up to the coefficient $C^k_{VV}$, but having some specific form in $S^1\times H^{d-1}$ that is different from flat space, which can be calculated using \eqref{eq:CHM}. Using the OPE \eqref{eq:OPE} the density matrix \eqref{eq:rhoV} has a simple expansion
\beq
\rho_V = e^{-\pi K} \sum_k C^k_{VV} B_k( \tau-\pi, \hat \tau-\pi) e^{-\pi K}.
\eeq

\subsection{Expansion of the noncommutative logarithm}
\label{sec:modularexpand}

Let us write the density matrix as
\beq
\rho_V = \rho_{\rm vac} + \delta \rho,
\eeq
where $\rho_{\rm vac} = e^{-2\pi K}$ is the vacuum density matrix. When $V$ is a primary, we have seen that
\beq
\label{eq:deltarho0}
\delta \rho = e^{-\pi K} \sum_{k\neq vac} C^k_{VV} B_k( \tau-\pi, \hat \tau-\pi)e^{-\pi K}.
\eeq
In any case, from \eqref{eq:rhoV}, the natural form of $\delta \rho$ is
\beq
\label{eq:deltarho}
\delta \rho = e^{-\pi K} \delta \tilde \rho e^{-\pi K}.
\eeq
In what follows, we set up a formal expansion for the modular Hamiltonian 
\beq
K_V = -\log \rho_V
\eeq
to all orders of $\delta \rho$. We will not be discussing convergence of this expansion, but instead we will be interested in the following situations.
We see that we stay close to $\rho_{\rm vac}$ in \eqref{eq:rhoV}, and therefore an expansion in $\delta \rho$ is useful, if the OPE \eqref{eq:OPE} is dominated by the identity contribution. One way to achieve this is to take a small subsystem size limit $\theta_0 \ll 1$, as discussed e.g. in \cite{Sarosi:2016oks,Sarosi:2016atx}. In this case, when one aims to extract the result up to a given power in the radius $\theta_0$ of the region, the series in $\delta \rho$ can be truncated. Another case of interest are states satisfying
\beq
C^k_{VV} \sim \langle V|\om_k|V\rangle \ll 1, \;\; k \neq \text{ vacuum}.
\eeq
Such states are natural in holography: they are perturbative supergravity excitations in the bulk around the AdS saddle, for which most of these coefficients are surpressed by $\frac{1}{N}$. The exceptions are the OPE coefficients of the single trace $V$ with its multitrace operators, such as $:VV:$, which are of order one. In this case, resumming these multitrace parts in our expansion would give the entanglement entropy to all orders in bulk perturbation theory. 

\vspace{4mm}
To derive the announced series representation, we will use the integral representation of the logarithm
\beq
-\log \rho = \int_0^\infty d\beta \left( \frac{1}{\beta+\rho}- \frac{1}{\beta+1}\right).
\eeq
Using this, we can write the modular Hamiltonian to all orders in $\delta \tilde \rho=e^{\pi K}\delta \rho e^{\pi K}$ as
\beq
\label{eq:modularexpansion}
K_V=2\pi K+\sum_{n=1}^\infty (-1)^n \delta K^{(n)},
\eeq
where
\beq
\delta K^{(n)} = \int_0^\infty d\beta \left[ (\beta + e^{-2\pi K})^{-1}e^{-\pi K}\delta \tilde \rho e^{-\pi K}\right]^n(\beta + e^{-2\pi K})^{-1}.
\eeq
Inserting an identity resolution in front of each $\delta \tilde \rho$ results in 
\beq
\label{eq:spectralresolution}
\delta K^{(n)} = \int d\omega_1 ... d\omega_{n+1} |\omega_1\rangle \langle \omega_1 |\delta \tilde \rho |\omega_2 \rangle ... \langle \omega_{n}|\delta \tilde \rho |\omega_{n+1}\rangle \langle \omega_{n+1}|  \left[  \int_0^\infty d\beta \frac{e^{\pi(\omega_1+\omega_{n+1})-2\pi \sum_{k=1}^{n+1} \omega_k}}{\prod_{k=1}^{n+1}(\beta+e^{-2\pi \omega_k})}  \right].
\eeq

The task now is to put the kernel in the rectangular parentheses into a nice form. Define the (invertible) linear redefinition of frequencies
\beq
\label{eq:newfrequencies}
a_i = \omega_i-\omega_{i+1}, \;\; i=1,...,n, \;\;\; b=\sum_{k=1}^{n+1} \omega_k.
\eeq
It is possible to show that the kernel is independent of $b$, i.e. when written as a function of $a_i$ and $b$ we have
\beq
\partial_b\left[  \int_0^\infty d\beta \frac{e^{\pi(\omega_1+\omega_{n+1})-2\pi \sum_{k=1}^{n+1} \omega_k}}{\prod_{k=1}^{n+1}(\beta+e^{-2\pi \omega_k})}  \right]=0.
\eeq
Indeed, noting that $\partial_b = \frac{1}{n+1} \sum_{k=1}^{n+1}\partial_{\omega_k}$ we see that $\partial_b f=0$ then is equivalent with demanding $f$ to be invariant under $\omega_i \rightarrow \omega_i + \alpha$. This is easily shown to hold by changing integration variable as $\beta=\tilde \beta e^{-2\pi \alpha}$.

It follows that we can represent the kernel with a Fourier integral involving only the $a_i$ variables
\beq
\label{eq:formalfourier}
\left[  \int_0^\infty d\beta \frac{e^{\pi(\omega_1+\omega_{n+1})-2\pi \sum_{k=1}^{n+1} \omega_k}}{\prod_{k=1}^{n+1}(\beta+e^{-2\pi \omega_k})}  \right] = \int_{-\infty}^{\infty} ds_1...ds_n e^{i \sum_{k=1}^n s_k a_k} \mathcal{K}_n(s_1,...,s_n).
\eeq
With the above Fourier representation, and using the expression \eqref{eq:newfrequencies} of $a_i$ with $\omega_k$ we can easily undo the spectral resolutions in expression \eqref{eq:spectralresolution} for $\delta K^{(n)}$ and arrive at the suggestive expression 
\beq
\label{eq:generalform}
\delta K^{(n)} = \int_{-\infty}^{\infty} ds_1...ds_n \mathcal{K}_n(s_1,...,s_n) \prod_{k=1}^n(e^{iKs_k}\delta \tilde \rho e^{-iKs_k}).
\eeq
This is the main result of the section, expressing the corrections to the modular Hamiltonian in terms of integrals along the vacuum modular flow. It applies to any quantum system, at this level it is just a convenient way of writing the Taylor expansion of the noncommutative log. We note that the kernel will have poles along the real line, therefore the above formula must be supplemented with a contour prescription. We continue by discussing the kernel and the contour prescription.

\subsection{The kernel}

Now we carry on to derive the explicit form of the modular kernel $\mathcal{K}_n(s_1,...,s_n)$. The inverse Fourier transform that we want to do is
\beq
\mathcal{K}_n(s_1,...,s_n) = \int \frac{ da_1...da_n}{(2\pi)^n} e^{-i \sum_{l=1}^n a_l s_l}\int_0^\infty d\beta \frac{e^{\pi(\omega_1+\omega_{n+1})-2\pi \sum_{k=1}^{n+1} \omega_k}}{\prod_{k=1}^{n+1}(\beta+ e^{-2\pi \omega_k})},
\eeq
with $\omega_i$ expressed via $a_i$ using \eqref{eq:newfrequencies}. The trick we use to decouple the multidimensional integral into a product of integrals is to multiply by $\delta(q) =\frac{1}{2\pi} \int_{-\infty}^\infty db e^{-iqb}$ and change integration variables back to $\omega_k$ using \eqref{eq:newfrequencies}. The resulting Jacobian is $n+1$. We will strip off the delta function at the end of the calculation. We have
\bea
\label{eq:interstep}
\delta(q)\mathcal{K}_n(s_1,...,s_n) &= \frac{n+1}{(2\pi)^{n+1}} \int_0^\infty d\beta \left[\int d\omega_1\frac{e^{-\pi \omega_1-i\omega_1(s_1+q)}}{\beta + e^{-2\pi \omega_1}} \right]
\left[\int d\omega_{n+1}\frac{e^{-\pi \omega_{n+1}+i\omega_{n+1}(s_n-q)}}{\beta + e^{-2\pi \omega_{n+1}}} \right] \\
&\times \prod_{k=2}^n \left[\int d\omega_{k}\frac{e^{-i\omega_{k}(s_k-s_{k-1}+q)-2\pi \omega_k}}{\beta + e^{-2\pi \omega_k}} \right]
\eea
To evaluate the integrals we need the formula
\beq
\label{eq:formula}
\int_{-\infty}^\infty d\omega \frac{e^{-\omega \xi}}{\beta+e^{-2\pi \omega}} = \frac{\beta^{\frac{\xi}{2\pi}-1}}{2 \sin \frac{\xi}{2}},
\eeq
valid for all $\xi \in \mathbb{C}^\times$ and $\text{Re}\beta \geq 0$. This formula is easily proved by closing the contour in either the upper or lower half plane (depending on the sign of $\text{Im}\xi$) and summing residues.

It is easy to check that when \eqref{eq:formula} is used, all the $s_i$ dependence in the powers of $\beta$ in the product in \eqref{eq:interstep} cancels and we end up with
\bea
\delta(q)\mathcal{K}_n(s_1,...,s_n) &= \frac{n+1}{(4\pi)^{n+1}} \frac{i^{n-1}}{\cosh \frac{s_1+q}{2} \cosh \frac{s_n-q}{2} \prod_{k=2}^n \sinh \frac{q+s_k-s_{k-1}}{2}} \\
&\times \int_0^\infty d\beta \beta^{-1+\frac{iq(n+1)}{2\pi}}.
\eea
Changing variables to $u=\log \beta$ in the integral we recognize $\frac{(2\pi)^2}{n+1} \delta(q)$. Therefore, the kernel reads as
\beq
\label{eq:modularkernel}
\mathcal{K}_n(s_1,...,s_n) = \frac{(2\pi)^2}{(4\pi)^{n+1}} \frac{i^{n-1}}{\cosh \frac{s_1}{2} \cosh \frac{s_n}{2} \prod_{k=2}^n \sinh \frac{s_{k}-s_{k-1}}{2}}.
\eeq
Some special cases:
\bea
\mathcal{K}_1(s_1) & = \frac{1}{(2 \cosh\frac{s_1}{2})^2}, \\
\mathcal{K}_2(s_1,s_2) & = \frac{1}{16\pi} \frac{i}{\cosh \frac{s_1}{2}\cosh \frac{s_2}{2}\sinh \frac{s_2-s_1}{2}}.
\eea
Note that Hermicity of $\delta K^{(2)}$ requires $\mathcal{K}_2(s_1,s_2)^* = \mathcal{K}_2(s_2,s_1)$, which is satisfied by this kernel. 

\subsection{The contour: part 1}

Now let us discuss the contour prescription in \eqref{eq:generalform}. 
When we put $\xi=ip+2\pi$, $p\in \mathbb{R}$ in \eqref{eq:formula} (we have in mind $p=s_{k}-s_{k-1}+q$ here, as in \eqref{eq:interstep}), we have the Fourier integral 
\beq
\int_{-\infty}^\infty d\omega \frac{e^{-2\pi \omega}}{\beta+e^{-2\pi \omega}}e^{-ip\omega} = -\frac{\beta^{\frac{ip}{2\pi}}}{2i \sinh \frac{p}{2}},
\eeq
which has the formal inverse
\beq
\label{eq:polepresc}
\int_{-\infty}^\infty \frac{ dp}{2\pi} e^{ip\omega} \frac{\beta^{\frac{ip}{2\pi}}}{2i \sinh \frac{p}{2}} = -\frac{e^{-2\pi \omega}}{\beta+e^{-2\pi \omega}}.
\eeq
However, we see that there is a pole at $p=0$ on the integration contour. It is easy to check that in order to reproduce the r.h.s. we need to go around this pole from \textit{above}, i.e. lift it as $p\rightarrow p+i\epsilon$ with $2\pi>\epsilon>0$. This is our contour prescription. It follows, that in the integrals of \eqref{eq:generalform} we must avoid poles\footnote{We note that while at this level, this prescription is merely needed for reproducing the expansion of the noncommutative log, when we later apply it to CFT correlators, we need a more refined contour prescription which properly fixes the order of Lorentzian operators in the correlators.} as $s_{k}-s_{k-1}\rightarrow s_{k}-s_{k-1}+i\epsilon$, $2\pi>\epsilon>0$.

\section{Entanglement entropy}
\label{ref:ee}

\subsection{Expansion for the entanglement entropy}
\label{ref:entropy}
We can write the entanglement entropy of the state $|V\rangle$ as
\beq
S_V =  \langle V|K_V|V\rangle,
\eeq
as we have picket the correct normalization for the density matrix. We are interested in the contribution above the first law part so we define
\beq
\label{eq:entropyexpansion0}
\delta S_V =\langle V|K_V|V\rangle-2\pi \langle V|K|V\rangle,
\eeq
where $K$ is the vacuum modular Hamiltonian. 
Then according to \eqref{eq:deltarho0}, \eqref{eq:modularexpansion} and \eqref{eq:generalform}
\bea
\label{eq:entropyexpansion1}
\delta S_V &=\sum_{n=1}^\infty   (-1)^n \langle V| \delta K^{(n)}|V\rangle,
\eea
where
\bea
\label{eq:entropyexpansion}
 &\langle V| \delta K^{(n)}|V\rangle = \\ &=\sum_{m_1,...,m_n\neq {\rm vac}}  \int_{-\infty}^{\infty} ds_1...ds_n \mathcal{K}_n(s_1,...,s_n) \frac{\langle V( \tau)V(\hat \tau)\left[\prod_{k=1}^n  C^{m_k}_{VV} B_{m_k}( \tau-\tau_{s_k},\hat \tau-\tau_{s_k}) \right]  \rangle_{\Sigma_1}}{ \langle V(\tau)V(\hat \tau) \rangle_{\Sigma_1} } \\
 &=\int_{-\infty}^{\infty} ds_1...ds_n \mathcal{K}_n(s_1,...,s_n)  \left\langle   \frac{V(\tau)V(\hat \tau)}{ \langle V(\tau)V(\hat \tau) \rangle_{\Sigma_1} }\left[\prod_{k=1}^n \left(  \frac{V(\tau-\tau_{s_k})V(\hat \tau-\tau_{s_k})}{\langle V( \tau-\tau_{s_k})V(\hat \tau-\tau_{s_k}) \rangle_{\Sigma_1}} -1\right)\right] \right\rangle_{\Sigma_1},
\eea
where $\tau_{s_k}=\pi-is_k$ and $\Sigma_1=S^1 \times H^{d-1}$. The kernel is given by \eqref{eq:modularkernel}. In order to interpret the multi-point function in the integrand as an analytically continued Schwinger function, the integral contours must be shifted from the real axis in a specific way which we will soon discuss. In summary, we have related the entanglement entropy to all orders in $C^{k}_{VV}$ to multipoint correlators of $V$ on $S^1 \times H^{d-1}$, without the need to perform any kind of analytic continuation. This is in principle calculable for any CFT, since the space $S^1 \times H^{d-1}$ for which the correlator have to be evaluated is conformal to flat space $\mathbb{R}^d$ via the map
\be
r= \f{\sinh u}{\cosh u +\cos \tau}, \qquad t= \f{\sin \tau}{\cosh u + \cos \tau},  \label{eq:hyptoflat}
\ee
for which
\be
ds^{2}_{S^{1} \times H^{d-1}} = \Omega^2 \left(  dt^2+ dr^2 +r^2 d\Omega_{d-2}^2 \right), \quad \Omega= \cosh u +\cos \tau .
\ee
The correlators in \eqref{eq:entropyexpansion} are particularly simple to map to flat space when $|V\rangle$ is a conformal primary, but the formula itself is applicable in principle to \textit{any state} $|V\rangle$ created by linear combinations of local operators.\footnote{When $V$ is not a conformal primary, we can still formally write $\delta \rho= e^{-\pi K}\left(\frac{V(\tau-\pi) V(\hat \tau-\pi)}{\langle V(\tau) V(\hat \tau) \rangle}-1\right)e^{-\pi K}$ and apply \eqref{eq:generalform} to get a formal series for the entanglement entropy.} Finally, we note that the piece of the first law part above the vacuum in \eqref{eq:entropyexpansion0} is calculable as 
\beq
2\pi\langle V|K|V\rangle-S_{\rm vac} =  2\pi \int_{H^{d-1}} d^{d-1}x \sqrt{g_{H^{d-1}}}\left(\frac{ \langle V(\tau) {T^\tau}_\tau (x)V(\hat \tau) \rangle_{\Sigma_1}}{\langle V(\tau) V(\hat \tau) \rangle_{\Sigma_1}} -\langle {T^\tau}_\tau (x)\rangle_{\Sigma_1}\right),
\eeq
where the correlators can be mapped to flat space using \eqref{eq:hyptoflat}. In flat space, the three point function involving the stress tensor can be found e.g. in \cite{Osborn:1993cr}. The effect of the last term is only to substract the conformal anomaly in the mapping of ${T^\tau}_\tau$ in even dimensions, which belongs to $S_{\rm vac}$  \cite{Casini:2011kv}. This ``first law part" is discussed to some extent e.g. in \cite{Wong:2013gua}, and we also see that it gives a linear contribution in the OPE coefficients $C^\om_{VV}$. Therefore, the complete vacuum substracted entanglement entropy $S_V-S_{\rm vac}$ is accessible.\footnote{The vacuum entropy $S_{\rm vac}$ is the thermal entropy of the space $S^1\times H^{d-1}$, which is rather hard to calculate in general for $d>2$ and depends heavily on the regularization scheme\cite{Casini:2011kv}. We are not discussing this in this paper. The ``first law substracted" entropy, $\delta S_V$ is on the other hand UV finite, as it can be written as a relative entropy between the excited state and the vacuum, see section \ref{sec:conv}.}

\subsection{The contour: part 2}
\label{sec:contourp2}

\begin{figure}[h!]
\centering
\includegraphics[width=0.4\textwidth]{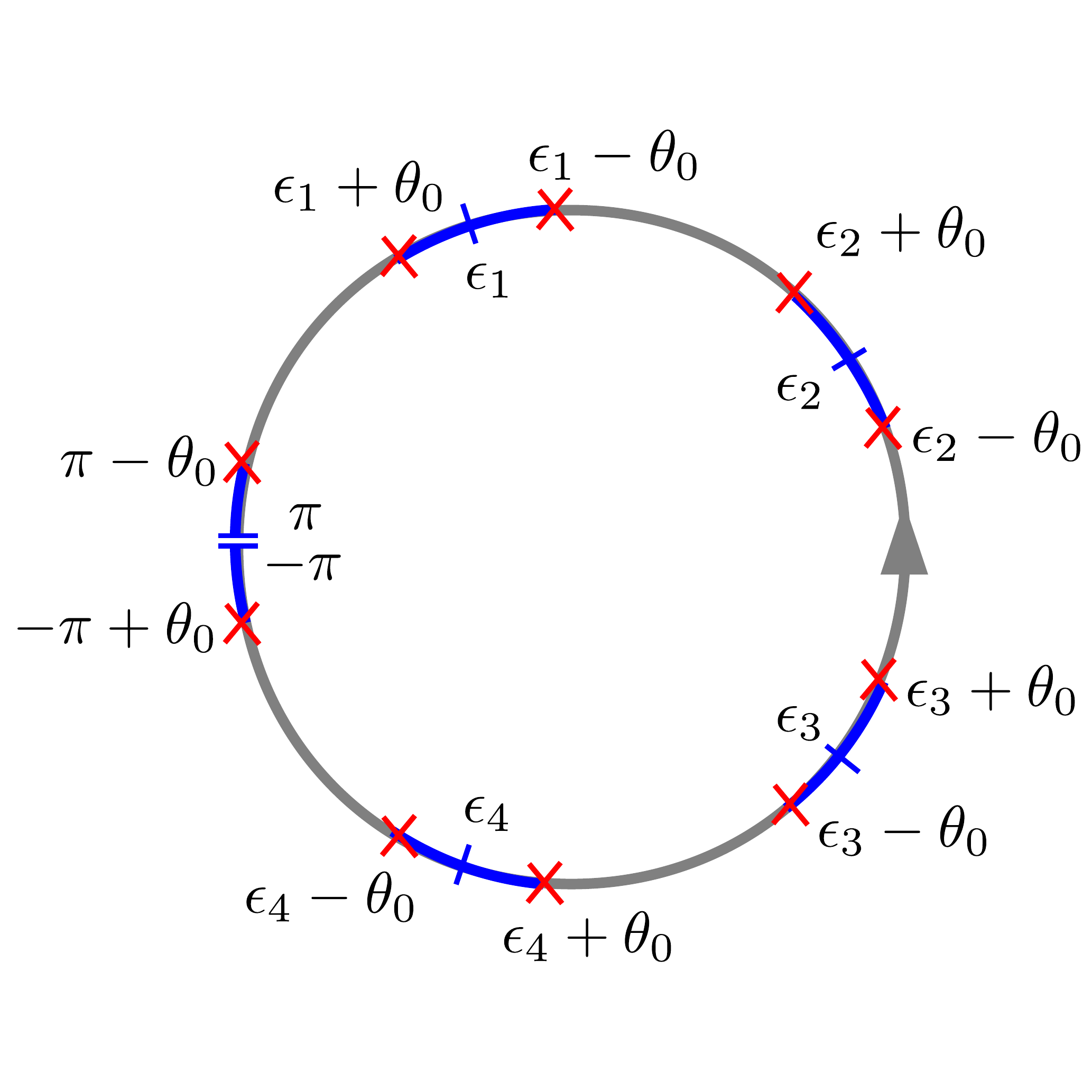} \hspace{1cm}
\includegraphics[width=0.4\textwidth]{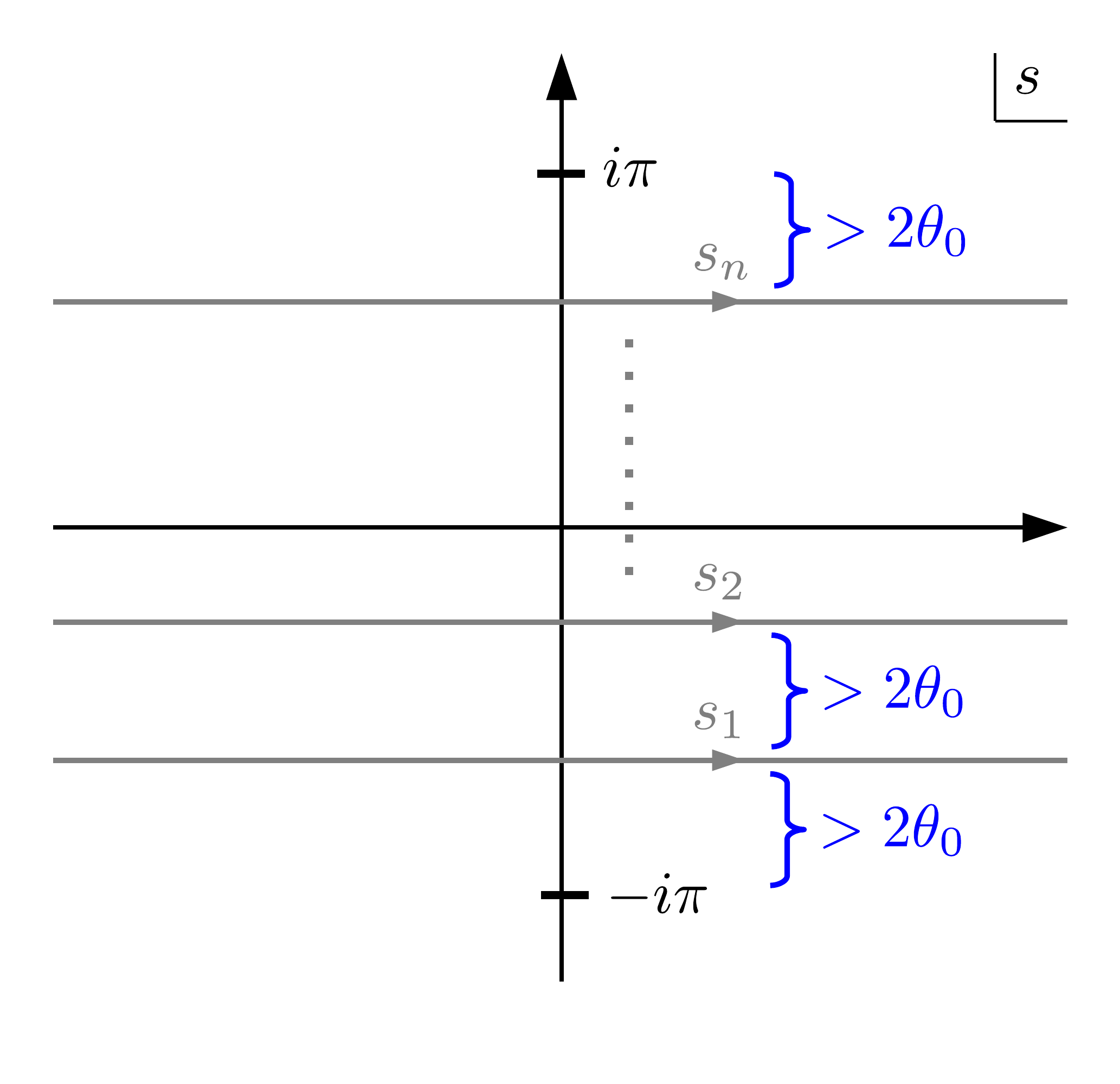}
\caption{Left: Ordering of operators in \eqref{eq:entropyexpansion}. The red crosses are the Euclidean projections of the operator insertions, and the right ordering is achieved from the Euclidean time ordering by picking the $s_k$ integral contours in \eqref{eq:entropyexpansion} accroding to \eqref{eq:contour}. Right: Illustration of the integration contours (grey) in \eqref{eq:entropyexpansion} prescribed in \eqref{eq:contour}.}
\label{fig:2}
\end{figure}

Formula \eqref{eq:generalform} requires a fixed ordering of operators, therefore to express the correlator in the integrand of \eqref{eq:entropyexpansion} with standard Schwinger functions, we must pick the $s_k$ integration contours so that they keep the operator ordering in \eqref{eq:entropyexpansion}. This is done by shifting the Lorentzian times
\beq
s_k \rightarrow s_k-i \epsilon_k,
\eeq
with operators inserted at larger Euclidean time being on the left.\footnote{Let us recall here an intuitive understanding of this prescription. The evolutions operator $e^{-iH\Delta t}$ for a complex time step $\Delta t$ is a bounded operator only if $\text{Im}\Delta t<0$. This is required for it to take normalizable states to normalizable states, and therefore to slice up the evolution into a convergent path integral. This shows that while the path integral contour can go in arbitrary direction in $\text{Re}t$, it can only go \textit{downwards} in $\text{Im}t$. Once $\text{Im}t$ decreased to a specific value, there is no turning back. It follows that operator insertions with a smaller imaginary value in time can only be on the later parts of the path integral contour.} This type of contour deformation is allowed in \eqref{eq:formalfourier} and hence \eqref{eq:generalform} provided
\beq
\pi>\epsilon_1>...>\epsilon_n>-\pi,
\eeq
since the kernel \eqref{eq:modularkernel} falls off quickly for large times and none of its poles are crossed for this range of $\epsilon_k$. Here, we took into account the way we were required to avoid singularities at $s_{k}-s_{k-1}=0$ in \eqref{eq:formalfourier}, see the discussion after \eqref{eq:polepresc}. To actually enforce the operator ordering in the correlator expressed with $V$ operators in \eqref{eq:entropyexpansion} we need to pick
\beq
\label{eq:contour}
 \pi-2\theta_0>\epsilon_k>-\pi+2\theta_0, \quad \text{and} \quad \epsilon_{k-1}-\epsilon_{k}>2\theta_0.
\eeq
The first requirement comes from ordering operators at $\hat \tau$ and $\tau-\tau_{s_{k}-i\epsilon_k}=-\theta_0+\epsilon_k+is_k$, while the second one from ordering $ \tau-\tau_{s_{k-1}-i\epsilon_{k-1}}=-\theta_0+\epsilon_{k-1}+is_{k-1}$ and $\hat \tau-\tau_{s_{k}-i\epsilon_k}=\theta_0+\epsilon_k+i s_k$, see Fig. \ref{fig:2} for an illustration of this.\footnote{We are grateful to Onkar Parrikar for pointing out a flaw in the contour prescription in a previous version of this manuscript.}  

\subsection{Question of convergence}
\label{sec:conv}

Notice that the $\epsilon$ prescription in \eqref{eq:contour} with straight integration contours as in the right of Fig. \ref{fig:2} is only possible when $\theta_0 < \pi /(n+1)$. When $\theta_0 > \pi /(n+1)$, the straight contours are no longer possible. This can be easily seen by examining Fig. \ref{fig:2}. One could imagine curly contours partially deformed over to the second sheet of the correlation function while still avoiding the branch point singularities in the appropriate way. However, we argue in Appendix \ref{app:3} that nevertheless each order becomes singular at least at $\theta_0=\pi/2$. We now briefly explain why this singularity cannot be physical and must disappear when the series is resummed. For this purpose, we introduce a control parameter $t$ for the expansion by considering the relative entropy
\begin{align}
\label{eq:relent}
S\big((1-t)\rho_{\rm vac} +t \rho_{V}|| \rho_{\rm vac}\big)&= -2\pi{\rm Tr} \left[ (\rho_{\rm vac} +t \delta \rho) K \right] -S(\rho_{\rm vac} +t\delta \rho),  \nonumber \\[+10 pt]
&= -\sum_{n=1}^{\infty} (-t)^n \text{Tr}\left[ (\rho_{\rm vac}+t\delta \rho)\delta K^{(n)}\right],
\end{align}
where $\delta \rho= \rho_{V}-\rho_{\rm vac}$ is just the same as \eqref{eq:deltarho} for primaries, and we have used (\ref{eq:entropyexpansion0},\ref{eq:entropyexpansion1}) to expand the entanglement entropy. The coefficient of $t^{n+1}$ in this expansion involves a combination of a finite number of at most $2n+2$ point functions with a number of integrals along the vacuum modular flow, similar to \eqref{eq:entropyexpansion}, each suffering from the divergence at $\theta_0=\pi/2$. However, the relative entropy itself is UV finite, and can only diverge when $\rho_{\rm vac}$ has some nontrivial kernel. By standard thermofield double decomposition, this only happens when we approach the total system, i.e. $\theta_0 \rightarrow \pi$, and the state becomes pure. Therefore, the relative entropy must be finite at $\theta_0=\pi/2$. Since the contribution of each order in the expansion blow up approaching this point, we must conclude that the series cannot converge for all $\theta_0$. In fact, if the $n$th term does blow up as $\theta_0 \rightarrow \pi/(n+1)$, then there is no convergence for any $\theta_0$. It is not so surprising that the $t$ expansion of \eqref{eq:relent} might have zero radius of convergence: for $t<0$ the first argument is not necessarily a valid density matrix so there is no reason to expect that \eqref{eq:relent} has a valid analytic continuation to any $t<0$.\footnote{It is also interesting to consider the properties of the expansion in $\theta_0$. The requirement $\theta_0 < \pi /(n+1)$ coming from \eqref{eq:contour} is reminiscent to the behavior of an asymptotic expansion in $\theta_0$. Indeed, for a given $n$ the smallest power appearing in \eqref{eq:entropyexpansion} is $\theta_0^{n\Delta}$, where $\Delta$ is the dimension of the lightest non-vacuum operator in the $V\times V$ OPE. Therefore, for a given $\theta_0$, it appears that we may only trust the series by expanding to order $\theta_0^k$, where $k<\Delta(\pi/\theta_0-1)$, which is a typical requirement for asymptotic series. Still, we do not claim that the $\theta_0$ expansion is asymptotic since collecting powers of $\theta_0$ involves a nontrivial reorganization of terms between the $n$ expansion and the OPE. Indeed, in the simplest examples, like when $V$ is a chiral vertex operator for the free boson in 2d, the relative entropy is known exactly \cite{Lashkari:2014yva} and has a non-zero radius of convergence in $\theta_0$.}

\subsection{The contour: part 3}
\label{sec:contourp3}

Studying convergence properties of this modular perturbation theory is an interesting question on its own right, but we will not pursue this in the present paper.  Instead, we will treat the operator ordering in \eqref{eq:entropyexpansion} by avoiding the resummation of the OPE blocks $B_k(\tau-\tau_{s_k},\hat \tau-\tau_{s_k})$ into correlators of $V$. This will turn out to be a natural thing to do when we want to interpret the appearing terms from a bulk point of view. In this case, the integrand in \eqref{eq:entropyexpansion} has the form
\beq
\sum_{m_1,...,m_n \neq vac}\left. C(\delta \tau,\partial_{\tau_1})...C(\delta \tau,\partial_{\tau_n}) \langle V( \tau) \om_{m_1}(\tau_1-\tau_{s_1})...\om_{m_n}(\tau_n-\tau_{s_n})  V(\hat \tau-2\pi) \rangle_{\Sigma_1}\right|_{\tau_{1}=...=\tau_{n}=\tau},
\eeq
where we have used \eqref{eq:OPB} and $\delta \tau=\hat \tau-\tau=2\theta_0$. Now we can get away with the simplified contour prescription
\beq
s_k \rightarrow s_k-i \tilde \epsilon_k, \quad \text{with} \quad  \pi>\tilde\epsilon_1>\tilde\epsilon_2>...>\tilde\epsilon_n>-\pi+2\theta_0,
\eeq
where there is no further constraint on $\tilde \epsilon_{k-1}-\tilde\epsilon_k>0$ so that we can use straight contours for the $s_k$ integrals for any $0<\theta_0<\pi$. In exchange, to have a convergent OPE for a given $m_k$ we need all other insertions to be further away from $\om_{m_k}(\tau-\tau_{s_k-i\tilde \epsilon_k})$ in Euclidean time than $\delta \tau=2\theta_0$. Taking this requirement into account for all $m_k$ leads to identical constraints on $\tilde \epsilon_k$ as in \eqref{eq:contour}. For many practical purposes, we can leave the OPE unsummed and work with the simplified contour, keeping in mind that we might be dealing with an asymptotic series. In the present paper we will always follow this approach.

\subsection{Quadratic order in the OPE coefficient}
\label{sec:quadratic}

Consider the decomposition of (\ref{eq:entropyexpansion0}-\ref{eq:entropyexpansion})
\beq
\label{eq:otherexpansion}
\delta S_V = \sum_{m=2}^\infty \delta S_V^{(m)},
\eeq
where $\delta S_V^{(m)}$ contains the product of $m$ nontrivial OPE coefficients $C^{k}_{VV}$. Note that $ \delta S_V^{(n)} \neq \langle V |\delta K^{(n)}|V\rangle$, due to the identity appearing in the OPE of the insertions realizing the expectation value. In this section we discuss the contributions to this at order $(C^{k}_{VV})^2$. By taking the external state OPE $VV=\langle V V\rangle(1+\sum_k C^k_{VV}B_k+...)$, we see that we only need to worry about $n=1$ and $n=2$ terms in \eqref{eq:entropyexpansion}
\bea
\label{eq:quadratic}
\delta S_V^{(2)} &= \sum_{k\neq vac}(C^k_{VV})^2 \Big[  -\int_{-\infty}^\infty ds \frac{\langle B_k(\tau-\tau_s,\hat \tau-\tau_s)B_k(\tau,\hat \tau)\rangle_{\Sigma_1}}{(2\cosh \frac{s}{2})^2}\\ & + \int ds_1 ds_2 \mathcal{K}_2(s_1,s_2) \langle B_k(\tau-\tau_{s_1},\hat \tau-\tau_{s_1})B_k(\tau-\tau_{s_2},\hat \tau-\tau_{s_2})\rangle_{\Sigma_1} \Big].
\eea
Here we have also used the orthogonality of the OPE blocks of different primaries in writing the sum. To simplify this expression, we show that the second term gives a contribution of the same form as the first term. To see this, first let us use translation invariance in the temporal direction of $S^1 \times H^{d-1}$ to rewrite the second term as
\beq
\int ds_1 ds_2 \mathcal{K}_2(s_1,s_2) \langle B_k(\tau,\hat \tau)B_k(\tau-\tau_{s_2}+\tau_{s_1},\hat \tau-\tau_{s_2}+\tau_{s_1},)\rangle_{\Sigma_1}.
\eeq
Now we exchange integration variables to $u=s_1+s_2$ and $v=s_2-s_1-i\pi$ (and deform the $v$ contour back to the real line, in accordance with our prescription on the contour) to write this as
\beq
\int dudv \frac{1}{16\pi} \frac{1}{\cosh\frac{v}{2}(\cosh\frac{u}{2}+i\sinh\frac{v}{2})}\langle B_k(\tau,\hat \tau)B_k(\tau-\tau_{v},\hat \tau-\tau_{v})\rangle_{\Sigma_1}.
\eeq
At this point, the integral over $u$ can be done and we end up with
\beq
\int dv \left( \frac{1}{2} \frac{1}{(2\cosh\frac{v}{2})^2} + \frac{v}{8\pi i(\cosh\frac{v}{2})^2}\right)\langle B_k(\tau,\hat \tau)B_k(\tau-\tau_{v},\hat \tau-\tau_{v})\rangle_{\Sigma_1}.
\eeq
Now it is possible to argue\footnote{Combine translational invariance with the KMS condition and use that the two OPE blocks commute due to nonzero (mod $2\pi$) Euclidean separation to show this.} that the correlator is symmetric under the Lorentzian time reflection $v\rightarrow -v$, therefore we can drop the term proportional to $v$. Putting back to \eqref{eq:quadratic} yields the simple expression
\beq
\label{eq:quadratic2}
\delta S_V^{(2)} = \sum_k (C^k_{VV})^2 \Big[  -\frac{1}{2}\int_{-\infty}^\infty ds \frac{\langle B_k(\tau-\tau_s,\hat \tau-\tau_s)B_k(\tau,\hat \tau)\rangle_{\Sigma_1}}{(2\cosh \frac{s}{2})^2} \Big].
\eeq
The two point function of the OPE block is just the four point conformal block $\mathcal{F}_{k}$ with intermediate primary $O_k$ 
\be
\mathcal{F}_{k}( x_{1},x_{2}, x_{3}, x_{4}) = \la B_{k} (x_{1}, x_{2}) B_{k} (x_{3}, x_{4}) \ra_{\Sigma_1} .
\ee
Therefore, the entanglement entropy at this order is given by an integral of the four point block on $S^1\times H^{d-1}$
\be
\delta S^{(2)}_V= -\sum_k (C^k_{VV})^2\int \f{ds}{8\cosh^2 \f{s}{2}}  \mathcal{F}_{k}(\tau, \hat{\tau} ,\tau-\tau_{s}, \hat{\tau}-\tau_{s}).
\ee

\subsection{Comparison with the replica trick}

The aim of this section is to confirm formula \eqref{eq:quadratic2} for the leading quadratic contribution, using the replica method. We remind the reader that our subregion $A$ is the cap $[0, \theta_{0}] \times S^{d-2}$. Via the replica trick, the entanglement  entropy is calculated from the R\'enyi entropy via
\be
S_{A} (\rho_{V}) = \lim_{n \rightarrow 1}\f{1}{1-n} {\rm Tr} \rho_{V}^{n},
\ee
where $\rho_{V}$ denotes the reduced density matrix of the excited state $|V \ra$ with respect to the region $A$.  The R\'enyi entropy is computed by a path integral on the $n$ sheeted cylinder glued together at the region $A$ with the boundary conditions specifying the excited state $|V \ra$. Furthermore, this $n$ sheeted geometry is conformally mapped to the manifold $\Sigma_{n}= S^{1}_{n} \times H^{d-1}$ with the metric\footnote{In the discussion below we will also use another set of coordinates $(Y_{I}, Y_{a}), a=1 \cdots d-1 $ on the hyperbolic space $H^{d-1}$, which is given by 
\be
(Y_{I}, Y_{a}) = (\cosh u, \sinh u \Omega_{a}), \quad -Y_{I}^2 + Y_{a}^2 =-1
\ee
 }
\be
ds^{2}_{\Sigma_{n}} = d\tau^2 + du^2+ \sinh^2 u\;  d\Omega_{d-2}^2, \quad \tau \sim \tau +2\pi n.
 \ee
In this frame, the R\'enyi entropy is computed by the $2n$ point correlation functions\footnote{We note in passing that the R\'enyi entropies of the Rindler density matrix \eqref{eq:rhoV} can be easily rewritten in this form, by interpreting $\text{Tr}(e^{-2\pi nK} ...)=\langle ...\rangle_{\Sigma_n}$. This shows that \eqref{eq:rhoV} has the right spectrum, giving some additional credibility to the somewhat heuristic path integral argument.} on the manifold $\Sigma_{n}$
\be
{\rm Tr} \rho_{V}^{n} = \f{\la \prod^{n-1}_{k=0} V( \tau_{k})V( \hat{\tau}_{k}) \rangle_{\Sigma_{n}} }{ \prod_{k=0}^{n-1}\la V( \tau_{k})V( \hat{\tau}_{k}) \ra_{\Sigma_{1}} } \equiv G_{n} \label{eq:corr}.
\ee
In the correlation function we suppressed the dependence on coordinates of $H^{d-1}$, as before. 
Since in the discussion below the excited state can be either static or time dependent, we leave the location of  the local operators unspecified.  However because of the replica symmetry of $\Sigma_{n}$, we have 
\be
(\tau_{k}, u_{k}) = (2\pi k + \tau, u), \quad  (\hat{\tau}_{k}, u_{k}) = (2\pi k + \hat{\tau}, \hat{u}) .
\ee
where $( \tau, u)$ and $( \hat{\tau}, \hat{u})$ denote the locations of the operators on the first sheet. For global states, we have
\be
(\tau, u) = (\pi -\theta_{0},0), \quad  (\hat \tau, u) = (\pi +\theta_{0},0),
\ee
as in \eqref{eq:globalinsertion}.
We note in passing that we can also think of more generic states like 
\be
\int_{S^{d-1}} d\Omega f (\Omega) V( \Omega,t) |0 \ra.
\ee
In this case the corresponding R\'enyi entropy is computed by an integral of the $2n$ point function. 


Now we would like to expand the correlation function (\ref{eq:corr})  in the channel $ \tau_{k} \rightarrow \hat{\tau}_{k}$.  
To do this, we use the OPE 
\be
\f{V(x_{1}) V(x_{2}) }{\la   V(x_{1}) V(x_{2}) \ra }=  \; C( x_{12} , \p_{2}) \om (x_{2}) +{\rm other \; primaries}
\ee
where $C( x_{12} , \p_{2})$ is the differential operator  defined in  (\ref{eq:OPB}).
Precisely speaking, to  expand the correlation function (\ref{eq:corr}) at $ n\neq 1$, we need to use the OPE in the presence of  a defect (or twist operators) which is a highly theory dependent object and we do not 
know its precise form. However, as is clear from the discussion below, to calculate the entanglement entropy we only need the usual OPE at $n=1$.     

By substituting the OPE into the correlation function $G_{2n}$, we can decompose it into $n$ pieces 
\be
 G_{2n} =\sum_{k=1}^{n} G_{2n}^{(k)},
\ee 
where $G_{2n}^{(k)} $ denotes the piece of the correlation function with $k$ non trivial internal operators.  Also, $G_{2n}^{(k)}$ is proportional to $(C_{VV\om})^{k}$, therefore the decomposition is regarded as the expansion with respect 
to the three point function $C_{VV\om}$, in a similar way as the expansion (\ref{eq:entropyexpansion0}-\ref{eq:entropyexpansion}). 

Let us concentrate on the $k=2$ term, where two nontrivial operators appear in the internal lines
\be
 \label{eq:2rep}
G_{2n}^{(2)}  =C( \delta \tau, \; \p_{a}) C( \delta \tau, \; \p_{b})\sum_{k,m=1, k <m}^{n-1} \left.\la \om (\tau_{m} +\tau_{a},Y_{a}) \om (\tau_{m} +\tau_{b},Y_{b}) \ra_{\Sigma_{n}} \right|_{\tau_{a}=\tau_{b}=0} .
\ee
Here, we used the fact that two  differential operators are identical, $ C( \delta \tau_{k}, \; \p)  = C( \delta \tau_{m}, \; \p) $ because of the $Z_{n}$ replica symmetry of $\Sigma_{n}$.
The sum in the right hand side of (\ref{eq:2rep})
\be 
\sigma_{n} =\frac{1}{2}\sum_{k,m=1}^{n-1} \la \om (\tau_{m} +\tau_{a}, Y_{a}) \om (\tau_{m} +\tau_{b}, Y_{b}) \ra_{\Sigma_{n}},
\ee
was discussed in \cite{Faulkner:2014jva}. The $n \rightarrow 1$ limit of it is given by 
\be
\lim_{n \rightarrow 1} \f{\sigma_{n}}{1-n} = -\int \f{ds}{8\cosh^2 \f{s}{2}}  \la \om (\tau_{s} +\tau_{a}, Y_{a}) \om ( \tau_{b}, Y_{b})  \ra_{\Sigma_{1}}, \quad \tau_{s} = \pi-is.
\ee  
Where the two point function on $\Sigma_{1}$ reads as 
\be 
\la \om (\tau_{a}, Y_{a}) \om ( \tau_{b}, Y_{b}) \ra_{\Sigma_{1}} = \f{1}{\left( -2Y_{a}Y_{b} -2\cos \tau_{ab} \right)^{2\Delta}}, \quad \tau_{ab} \equiv \tau_{a} -\tau_{b}.
\ee

\noindent
Combining the above results, the entanglement entropy at this order is given by 
\be 
S^{(2)} =-C( \delta \tau, \; \p_{a}) C( \delta \tau, \; \p_{b}) \left. \int \f{ds}{8\cosh^2 \f{s}{2}}  \la \om (\tau_{s} +\tau_{a}, Y_{a}) \om ( \tau_{b}, Y_{b})  \ra_{\Sigma_{1}} \right|_{\tau_{a}=\tau_{b}=0}.
\ee
By using the definition of OPE block \eqref{eq:OPB}, we get 
\begin{align}
S^{(2)}&=-C( \delta \tau, \; \p_{a}) C( \delta \tau, \; \p_{b})  \int \f{ds}{8\cosh^2 \f{s}{2}} \left. \la \om (\tau_{s} +\tau_{a}, Y_{a}) \om ( \tau_{b}, Y_{b}) \ra_{\Sigma_{1}} \right|_{\tau_{a}=\tau_{b}=0}  \nonumber  \\[+10pt] 
&=- C^\om_{VV  }C( \delta \tau, \; \p_{a})\int \f{ds}{8\cosh^2 \f{s}{2}} \left. \la \om (\tau_{s} +\tau_{a}) B_{\om} ( \tau, \hat{\tau}) \ra_{\Sigma_{1}} \right|_{\tau_{a}=\tau_{b}=0}  \nonumber  \\[+10pt] 
&=- C^\om_{VV  }C( \delta \tau, \; \p_{a})\int \f{ds}{8\cosh^2 \f{s}{2}} \left. \la \om (\tau_{a}) B_{\om} ( \tau-\tau_{s}, \hat{\tau}-\tau_{s}) \ra_{\Sigma_{1}} \right|_{\tau_{a}=\tau_{b}=0}  \nonumber  \\[+10pt] 
&=-(C^\om_{VV  })^2\int \f{ds}{8\cosh^2 \f{s}{2}}  \la B_{\om} (\tau, \hat{\tau}) B_{\om} ( \tau-\tau_{s}, \hat{\tau}-\tau_{s}) \ra_{\Sigma_{1}},  \nonumber   
\end{align}
which is the same as \eqref{eq:quadratic2}.

\subsection{Vacuum Fisher information}
\label{sec:Fisher}


Let us consider the parameter dependent relative entropy given in \eqref{eq:relent}. The derivatives of this relative entropy with respect to $t$ are always well defined irrespective of the values of the OPE coefficients $C^\om_{VV }$
\be 
\f{d^{k}}{dt^{k}}S\big((1-t)\rho_{\rm vac} +t \rho_{V}|| \rho_{\rm vac}\big).
\ee
When $k=1$ this is identically vanishing because of the entanglement first law \cite{Bhattacharya:2012mi}. When $k=2$, this quantity at $t=0$ is known as quantum Fisher information, which can be viewed as a metric on the space of the reduced density matrices. It is simply given by
\bea
\f{d^{2}}{dt^{2}}S\big((1-t)\rho_{\rm vac} +t \rho_{V}|| \rho_{\rm vac}\big)|_{t=0} &= 2\text{Tr}(\delta \rho \delta K^{(1)}-\rho_{\rm vac}\delta K^{(2)})\\
&=\sum_{k \neq vac}(C^k_{VV})^2 \int_{-\infty}^\infty \f{ds}{8\cosh^2 \f{s}{2}}  \mathcal{F}_{k}(\tau, \hat{\tau} ,\tau-\tau_{s}, \hat{\tau}-\tau_{s}),
\eea
where we have used the results of section \ref{sec:quadratic} to cast the two terms into the same form. This is an exact expression for the quantum Fisher information, applicable when $\theta_0 <\pi/2$, i.e. half the size of the total system. This restriction comes from the contour prescription discussed in section \ref{sec:contourp2}. The discussion in Appendix \ref{app:3} suggests that this Fisher information diverges as we take $\theta_0 \rightarrow \pi/2$. It is a general result, that the quantum Fisher information is the classical Fisher information for the probability distribution associated by the state to some optimal observable \cite{petz2011introduction}. Then this divergence roughly means that we can learn a lot about the value of $t$ by measuring this optimal observable. It would be interesting to understand why is this happening at half the size of the total system.

\subsection{Cubic order in the OPE coefficient}
\label{sec:cubic}

Let us pick up the $(C^\om_{VV})^3$ terms in \eqref{eq:entropyexpansion}, which now requires $n=2$ and $n=3$ components, depending on which part of the $VV$ OPE we take:
\bea
\delta S^{(3)}_V &=(C^{\om}_{VV})^3  \Big[ \int_{-\infty}^{\infty} ds_1ds_2 \mathcal{K}_2(s_1,s_2) \\ &\times \langle  B_\om(\tau-\tau_{s_1},\hat \tau-\tau_{s_1}) B_\om(\tau-\tau_{s_2},\hat \tau-\tau_{s_2})  B_\om(\tau,\hat \tau)  \rangle_{\Sigma_1} \\ &-
\int_{-\infty}^{\infty} ds_1ds_2ds_3 \mathcal{K}_3(s_1,s_2,s_3) \\ &\times \langle   B_\om(\tau-\tau_{s_1},\hat \tau-\tau_{s_1})B_\om(\tau-\tau_{s_2},\hat \tau-\tau_{s_2}) B_\om(\tau-\tau_{s_3},\hat \tau-\tau_{s_3})   \rangle_{\Sigma_1} 
\Big]\\
&+ \cdots .
\eea
The dots represent contributions from other operators to the $m=3$ term in \eqref{eq:otherexpansion}. In particular, the three point function of OPE blocks is nonzero even for three different primaries, but for notational simplicity, here we are only going to focus on the $(C^\om_{VV})^3$ terms for a given primary $\om$ in the $V\times V$ OPE.
We again simplify the above expression by casting the second term to the same form as the first term. This is done in the same way as for the quadratic piece: first we shift all arguments by $\tau_{s_2}$ and then we introduce new integration variables
\bea
v_1=s_3-s_2, && v_2=s_1-s_2, && u=s_1+s_2+s_3.
\eea
The correlator is independent of $u$ and this integral can be done rather easily\footnote{Substitute $m=e^{u/3}$ to cast it into rational form.} 
\bea
\int_{-\infty}^{\infty} ds_1ds_2ds_3 \mathcal{K}_3(s_1,s_2,s_3)  &\langle   B_\om(\tau-\tau_{s_1},\hat \tau-\tau_{s_1})B_\om(\tau-\tau_{s_2},\hat \tau-\tau_{s_2}) B_\om(\tau-\tau_{s_3},\hat \tau-\tau_{s_3})   \rangle_{\Sigma_1} 
\\
&=\int_{-\infty}^\infty dv_1 dv_2 \frac{1}{32\pi^2}\frac{v_2-v_1}{\sinh\frac{v_1}{2}\sinh\frac{v_2}{2}\sinh\frac{v_2-v_1}{2}} \\ &\times \langle  B_\om(\tau+i v_2,\hat \tau+i v_2) B_\om(\tau,\hat \tau)  B_\om(\tau+i v_1,\hat \tau+i v_1)  \rangle_{\Sigma_1}.
\eea
Now, we deform the contours as $v_2\rightarrow v_2-i\epsilon_2$ and $v_1 \rightarrow v_1-i\epsilon_1$, with $\pi>\epsilon_2>0>\epsilon_1>-\pi$ to facilitate the operator ordering. After this, we move the first operator to the back using the KMS relation (this amounts to a $-2\pi$ shift in its arguments) and then choose $\epsilon_1=-\pi+\delta$, $\epsilon_2=\pi-\delta$, $\delta>0$, $\delta \rightarrow 0$. The result is
\bea
\int_{-\infty}^{\infty} ds_1ds_2ds_3 \mathcal{K}_3(s_1,s_2,s_3) & \langle   B_\om(\tau-\tau_{s_1},\hat \tau-\tau_{s_1})B_\om(\tau-\tau_{s_2},\hat \tau-\tau_{s_2}) B_\om(\tau-\tau_{s_3},\hat \tau-\tau_{s_3})   \rangle_{\Sigma_1} 
\\
&=\int_{-\infty}^\infty dv_1 dv_2 \mathcal{K}_2(v_1,v_2)\left(1 +i\frac{v_2-v_1}{2\pi}\right)\\ &\times \langle  B_\om(\tau,\hat \tau) B_\om(\tau-\tau_{v_1},\hat \tau-\tau_{v_1})  B_\om(\tau-\tau_{v_2},\hat \tau-\tau_{v_2})  \rangle_{\Sigma_1},
\eea
where we keep in mind that branch cuts along real $v_2-v_1$ must be avoided as $v_2-v_1 \rightarrow v_2-v_1+i\delta$, $\delta>0$.
The upshot of all this is that we can write the complete cubic term as
\bea
\label{eq:cubicee}
\delta S^{(3)}_V &= -(C^{\om}_{VV})^3   \int_{-\infty}^{\infty} ds_1ds_2 \mathcal{K}_2(s_1,s_2) \\ &\times \frac{i(s_2-s_1)}{2\pi} \langle  B_\om(\tau-\tau_{s_1},\hat \tau-\tau_{s_1}) B_\om(\tau-\tau_{s_2},\hat \tau-\tau_{s_2})  B_\om(\tau,\hat \tau)  \rangle_{\Sigma_1}\\
&+\cdots .
\eea
Finally, we note that there are always two similar terms contributing to a given $\delta S_V^{(m)}$ in \eqref{eq:otherexpansion}. The second one can be always simplified by performing the integral over $u=\sum_i s_i$, of which the correlator is independent.

\section{Holographic interpretation}
\label{sec:holo}

\subsection{Entanglement entropy and canonical energy}

We saw in section \ref{sec:quadratic} that at seconder order in the expansion in terms of $C_{VV\om}$, the entanglement entropy is given by the integral of  the four point conformal block along the vacuum modular flow 
\begin{align}
\delta S^{(2)}_V&=-(C^\om_{VV})^2 \int^{\infty}_{-\infty} \f{ds}{8\cosh^{2} \f{s}{2}} \mathcal{F}_{\om}( \tau,  \hat{\tau}, \tau-\tau_{s}, \hat{\tau}-\tau_{s}) , \quad \tau_{s}=\pi-is. \nonumber \\ 
&= -C(\delta \tau, \p_{a})  C(\delta \tau, \p_{b})\int^{\infty}_{-\infty} \left. \f{ds}{8\cosh^{2} \f{s}{2}} \la \om (\tau_{a}+\tau_{s},Y_{a}) \om (\tau_{b},Y_{b}) \ra_{\Sigma_{1}}  \right|_{(\tau_{a},Y_{a})=(\tau_{b},Y_{b})=(0,0)},
\end{align}
where $C(\delta \tau, \p_{a})$ is the differential operator summing up the descendants of the primary $\om (\tau_{a}, Y_{a})$,
\be
V(\tau) V(\hat{\tau}) = \la V(\tau) V (\hat{\tau}) \ra_{\Sigma_{1}}  C(\delta \tau, \p_{a}) \om (\tau_{a}, Y_{a}) + {\rm other \;  primaries}
\ee

We would like to provide a holographic interpretation of this formula. The discussion of this section is a slight modification of the one in \cite{Faulkner:2014jva}, where the change of the vacuum entanglement entropy under deformations of the CFT, and its holographic interpretation was discussed. In particular, in \cite{Faulkner:2014jva} it was shown 
that the integral of the CFT two point function along the vacuum modular flow can be written in terms of the integral of bulk-to-boundary propagators on the AdS hyperbolic black hole \footnote{In this section we employ the following normalization of the CFT two point function 
\be 
\la \om (\tau_{a}, Y_{a}) \om ( \tau_{b}, Y_{b}) \ra_{\Sigma_{1}} = \f{c_{\Delta}}{\left( -2Y_{a}Y_{b} -2\cos \tau_{ab} \right)^{2\Delta}}, \quad \tau_{ab} \equiv \tau_{a} -\tau_{b}, \quad c_{\Delta} =\f{2(\Delta-\f{d}{2} )\Gamma(\Delta)}{\pi^{\f{d}{2}}\Gamma(\Delta-\f{d}{2})},
\ee
so that the kinetic term of the dual bulk scalar field $\phi$ Lagrangian is properly normalized.}
 
\begin{align} 
I&= \int^{\infty}_{-\infty} \f{ds}{4\cosh^{2} \f{s}{2}} \la \om (\tau_{a}+\tau_{s},Y_{a}) \om (\tau_{b},Y_{b}) \ra_{\Sigma_{1}}  \nonumber \\[+10pt]
&=\f{\pi c_{\Delta}^2}{\left(\Delta -\f{d}{2} \right)^2} \int d l_{B} l_{B} \int dY_{B} \f{\p}{\p l_{B}} \left[ \f{1}{(-2Y_{a}Y_{B}-l_{B} e^{-i \tau_{a}})^{\Delta}}\right] 
\f{\p}{\p l_{B}} \left[ \f{1}{(-2Y_{b}Y_{B}-l_{B} e^{-i \tau_{b}})^{\Delta}}\right] ,
\end{align}
where $(l_{B}, Y_{B})$ denote the coordinates of the future horizon of the hyperbolic black hole. We follow the notation of \cite{Faulkner:2014jva} for the coordinates of the hyperbolic black hole, which we review in Appendix \ref{section:hb}. 
Notice that each quantity in the bracket $[ \cdots ]$ is nothing but the  bulk to boundary propagator\footnote{The easiest way to see this is to write the expectation value of the bulk field $\phi$ with source $\phi_0$ as $\langle\phi(X)\rangle_{\phi_0}=\int dx G_{\rm bulk \rightarrow bndy}(X|x)\phi_0(x)$, take a variation with respect to the source, and set it to zero.}
\be
\la \phi (l_{B},Y_{B}) \om (\tau_{a}, Y_{a}) \ra_{\Sigma_{1}}  = \f{a_{\Delta}}{(-2Y_{a}Y_{B}-l_{B} e^{-i \tau_{a}})^{\Delta}}, \quad a_{\Delta} =\f{c_{\Delta}}{2(\Delta-\f{d}{2})}.
\ee  
where $\phi$ is the bulk extension (via the HKLL procedure \cite{Hamilton:2006az}) of the primary $\om$.
This means that 
\begin{align}
C(\delta \tau, \p_{a}) \left[\f{a_{\Delta}}{(-2Y_{a}Y_{B}-l_{B} e^{-i \tau_{a}})^{\Delta}} \right] &=C(\delta \tau, \p_{a}) \la \phi (l_{B},Y_{B}) \om (\tau_{a}, Y_{a}) \ra_{\Sigma_{1}}   \nonumber \\[+10 pt]
&= \f{\la \phi (l_{B},Y_{B})  V( \tau) V(\hat{\tau}) \ra_{\Sigma_{1}} }{ \la V(\tau ) V(\hat{\tau} ) \ra_{\Sigma_{1}} } \nonumber \\[+10 pt] 
&\equiv \la\phi (l_{B},Y_{B}) \ra_{V}.
\end{align}
Notice that the expectation value $ \la  \phi (l_{B},Y_{B})\ra_{V}$ defined in the last line is regarded as the expectation value of the bulk scalar operator in the CFT excited state $| V \ra$, and that it satisfies the equation of motion of a bulk scalar field with the boundary condition for normalizable modes on the Lorenzian section of the black hole.
By combining  these results  we  have 
\be
\delta S^{(2)}_V = -2\pi\int d l_{B} l_{B} \int dY_{B} (\p_{l_{B}} \la\phi  \ra_{V} )^2\label{eq:int2}
\ee
The integrand is nothing but the null component of the stress tensor of the bulk scalar field, 
\be
T_{ab} = \p_{a} \phi \p_{b}\phi -g_{ab} \left (\f{1}{2} (\p \phi)^2+\f{1}{2} m^2 \phi^2\right).
\ee 
where $g_{ab}$ denotes the metric of the (unperturbed) hyperbolic black hole. 

Next, let us rewrite the integral \eqref{eq:int2} in global AdS
\be 
ds^2 =-(r^2+1)dt^2 +\f{dr^2}{r^2+1} +r^2 d\Omega_{d-2}^2.
\ee
We need to do this because we have originally defined our CFT on the cylinder $ \mathbb{R} \times S^{d-1}$, which is the boundary of global AdS. 
The bifurcation surface of the hyperbolic black hole corresponds to the RT surface of the region $A$ in global AdS, whose area computes the vacuum entanglement entropy.  Because of the conservation law of the stress tensor, $ \nabla^{a} T_{ab} =0$, one can deform the domain of integration, and  write \eqref{eq:int2} as the  integral on the region $\Sigma$ at $t=0$ which is enclosed by the  RT surface and the boundary subsystem $A$.\footnote{More precisely, the horizon integral is divided into two pieces by the deformation. One is the integral on $\Sigma$ which we have mentioned, and the other piece is the integral on the future part of the causal diamond of the boundary subsystem $A$. However the latter piece vanishes since we are dealing with the normalizable mode  of the bulk field.} Therefore,
\be
\label{eq:entropycanonicalenergy}
\delta S^{(2)}_V = -2\pi \int_{\Sigma} d \Sigma^{a} \xi^{b}  T_{ab} (\la \phi \ra_{V}),
\ee
where $T_{ab} (\la \phi \ra_{V}) $ is the stress energy tensor of the (classical) bulk scalar field profile $\la \phi \ra_{V}$, and $\xi_{b} $ is the bulk Killing field generating the casual cone of region $\Sigma$, called the entanglement wedge. 

In \cite{Lin:2014hva,Lashkari:2015hha,Nozaki:2013vta} it was argued using the Ryu-Takayanagi formula that in a CFT with a gravity dual, the quantum Fisher information (or equivalently the second order change of entanglement entropy in the one point function $\la \om \ra$) is interpreted as the canonical energy in the gravity side. Our derivation here is not only proving this statement, but also  telling us that  the Fisher information of ${\it any}$ CFT can be interpreted as a bulk canonical energy in AdS\footnote{Here we only focused on the contribution of scalar intermediate operators. We expect that similar relations hold for operators with spin.}, since in deriving the statement we only used the conformal symmetry of the CFT.

\subsection{Modular Hamiltonian}

The leading order corrections in the OPE coefficients to the modular Hamiltonian coming from \eqref{eq:generalform} can be written as
\beq
\label{eq:modularleading}
K_V = 2\pi K- \sum_{k\neq vac} C^k_{VV}\int_{-\infty}^\infty ds\frac{B_k(\tau-\tau_s,\hat \tau-\tau_s)}{(2\cosh\frac{s}{2})^2} + \cdots.
\eeq
The dots, here and below, denote contributions coming with higher powers of $C^k_{VV}$.
Notice that this can be obtained directly from the entanglement entropy \eqref{eq:quadratic2} by applying the ``first law trick", i.e. taking a variation in the state $V$ and using the first law $\delta S_V =  \text{Tr}(\delta\rho_V K_V)$. To see this, use the differential operator representation of the OPE blocks as in \eqref{eq:OPB} and write the one point functions in the expansion as expectation values in the state $\rho_V$. This trick is not guaranteed to work, as we do not know the form of the entanglement entropy for general variations of $\rho_V$, but in the present case it can be explicitly checked that it gives the right result. Now applying the trick to \eqref{eq:entropycanonicalenergy} gives a formula for the leading correction in terms of the HKLL reconstructed bulk field
\beq
\label{eq:bulkmodular}
K_V=2\pi K- 2 \pi \int_\Sigma d \Sigma^{a} \xi^{b}  \tilde T_{ab} (\la \phi \ra_{V}, \phi)+\cdots,
\eeq
where
\bea
\tilde T_{ab} (\la \phi \ra_{V}, \phi) &= \int \frac{\delta T_{ab} (\la \phi \ra_{V})}{\delta \la \phi \ra_{V}}\phi \\
&=2\p_{a} \la \phi \ra_{V} \p_{b}\phi -g_{ab} \left ( \p \la \phi \ra_{V}\p \phi+m^2 \la \phi \ra_{V}\phi\right),
\eea
i.e. twice the free field stress tensor with one of the expectation values $\la \phi \ra_{V}$ in the bilinear replaced by the operator $\phi$. Comparing with \eqref{eq:modularleading} this also results in the expression
\beq
C^\om_{VV}\int_{-\infty}^\infty ds\frac{B_\om(\tau-\tau_s,\hat \tau-\tau_s)}{(2\cosh\frac{s}{2})^2}=2\pi \int_\Sigma d \Sigma^{a} \xi^{b}  \tilde T_{ab} (\la \phi \ra_{V}, \phi),
\eeq
which is an identity involving the boundary OPE block and the integral of the bulk field on a codimension 1 bulk surface with some smearing function. This is in spirit similar to the OPE block/surface operator correspondence of \cite{Czech:2016xec}, and both formulas are special cases of the general relation derived recently in \cite{Faulkner:2017vdd}. 

Another suggestive way of rewriting \eqref{eq:bulkmodular} is
\beq
\label{eq:suggestive}
K_V = 2\pi \left( K- \int_\Sigma d \Sigma^{a} \xi^{b}  T_{ab} (\phi) \right) +2\pi \int_\Sigma d \Sigma^{a} \xi^{b}  T_{ab} (\phi-\la \phi \ra_{V}) + \delta S^{(2)}_V +\cdots.
\eeq
The expression $2\pi \int_\Sigma d \Sigma^{a} \xi^{b}  T_{ab} (\phi)$ is just the vacuum modular Hamiltonian of a free scalar $\phi$ associated to the bulk region $\Sigma$ in AdS, see e.g. \cite{Jafferis:2015del}. From the CFT point of view, it is identical to the effective generalized free field modular Hamiltonian introduced in \cite{Faulkner:2017vdd}.
The piece in the first big parentheses of \eqref{eq:suggestive} can be interpreted as the perturbative area operator $\frac{A}{4G}$ around empty AdS, following Hollands-Wald type arguments \cite{Hollands:2012sf,Lashkari:2015hha,Jafferis:2015del,Lashkari:2016idm}. We interpret the rest as $K_V^{\rm bulk}$, i.e. the bulk modular Hamiltonian associated to the state $|V\rangle$. Notice that when $\phi$ is a free field in $AdS$ (i.e. the dual operator is a generalized free field), we can actually interpret this as the modular Hamiltonian of a coherent state.\footnote{We thank Aitor Lewkowycz for pointing this out to us.} This is because
\beq
\label{eq:bulkmod}
2\pi \int_\Sigma d \Sigma^{a} \xi^{b}  T_{ab} (\phi-\la \phi \ra_{V}) = U_V \Big( 2\pi \int_\Sigma d \Sigma^{a} \xi^{b}  T_{ab} (\phi) \Big) U_V^{-1},
\eeq 
where $U_V=e^{i\int (\langle \phi \rangle_V \pi-\langle \pi \rangle_V \phi)}$ is the Weyl operator ($\pi$ here is canonical momentum) creating a coherent state with classical profile $\langle \phi \rangle_V$, $\langle \pi \rangle_V$ from the vacuum. Then, \eqref{eq:bulkmod} is just the modular Hamiltonian of the bulk QFT state $U_V|0\rangle$. This is true regardless of the fact that $|V\rangle$ is an energy eigenstate in the bulk. This difference is only reflected in the extra change in entanglement entropy, $\delta S^{(2)}_V$, which would be absent for a real bulk coherent state. This way, \eqref{eq:suggestive} is a special version of the conjecture put forward in \cite{Jafferis:2015del} which holds universally for \textit{any} CFT.

\subsection{Relative entropy}

It is straightforward to derive a suggestive expression for the relative entropy at quadratic order in the one point functions $C^\om_{VV}$ and $C^\om_{WW}$ of two light states $|V\rangle$ and $|W\rangle$, using \eqref{eq:bulkmodular}. The relative entropy at this order is symmetric, and has the bulk expression
\bea
S(\rho_V||\rho_W) &=  \Delta \langle K_W \rangle - \Delta S\\
&=2\pi \int_\Sigma d \Sigma^{a} \xi^{b}  T_{ab} (\la \phi \ra_{V}-\la \phi \ra_{W})+\cdots,
\eea
i.e. it is the canonical energy of the field configuration $\la \phi \ra_{V}-\la \phi \ra_{W}$. This is also proportional to the exact distance between the states in the vacuum Fisher information metric, as was first pointed out in \cite{Lashkari:2015hha} for holographic theories. Notice, however, that we have derived this formula without assuming the Ryu-Takayanagi formula, or other holographic properties of the CFT.

\section{Some explicit results in the small subsystem expansion}
\label{sec:smallsub}

To get our feet wet with applying the new method to calculate entanglement entropy, here we would like to compare to some of the results of \cite{Sarosi:2016atx} in the small subsystem size limit. As mentioned before, when one is interested in the expansion of the entanglement entropy up to some given order in the subsystem size $\theta_0$, the series (\ref{eq:entropyexpansion0}-\ref{eq:entropyexpansion}) can be truncated. This is because the smallest power appearing in the $n$th term, \eqref{eq:entropyexpansion}, is $\theta_0^{n\Delta}$, where $\Delta$ is the dimension of the lightest nonvacuum operator $O$ in the $V \times V$ OPE. Below, we will pick up the powers $\theta_0^{2\Delta}$ and $\theta_0^{3\Delta}$.

\subsection{Bioperator exchange}

The aim is to pick up the $\theta_0^{2\Delta}$ contribution from $\delta S_V$ as $\theta_0\rightarrow 0$. In the small subsystem limit we have
\beq
B_O( \tau, \hat \tau) \approx (2\theta_0)^{\Delta} O(\hat \tau) + \cdots.
\eeq
Using this in \eqref{eq:quadratic2} gives
\beq
\delta S_V^{(2)} = (C^O_{VV})^2 (2\theta_0)^{2\Delta}\left[  -\frac{1}{2}\int_{-\infty}^\infty ds \frac{\langle O(\hat \tau-\tau_s)O(\hat \tau)\rangle_{\Sigma_1}}{(2\cosh \frac{s}{2})^2} \right].
\eeq
Using \eqref{eq:int1} we have
\bea
\int_{-\infty}^\infty ds \frac{\langle O(\hat \tau-\tau_s)O(\hat \tau)\rangle_{\Sigma_1}}{(2\cosh \frac{s}{2})^2} 
&= \int_{-\infty}^\infty ds \frac{1}{(2\cosh \frac{s}{2})^{2+2\Delta}}\\
& = \frac{\sqrt{\pi}\Gamma(\Delta+1)}{2^{1+2\Delta}\Gamma(\Delta+\frac{3}{2})} ,
\eea
where we have used the map \eqref{eq:hyptoflat} to calculate
\beq
\langle O(\tau)O(0)\rangle_{\Sigma_1}=\f{1}{\big((2 \sin \frac{\tau}{2})^2\big)^{\Delta_O}}
\eeq
This agrees with the result of \cite{Sarosi:2016oks,Sarosi:2016atx}.

\subsection{Trioperator exchange}

\begin{figure}[h!]
\centering
\includegraphics[width=0.7\textwidth]{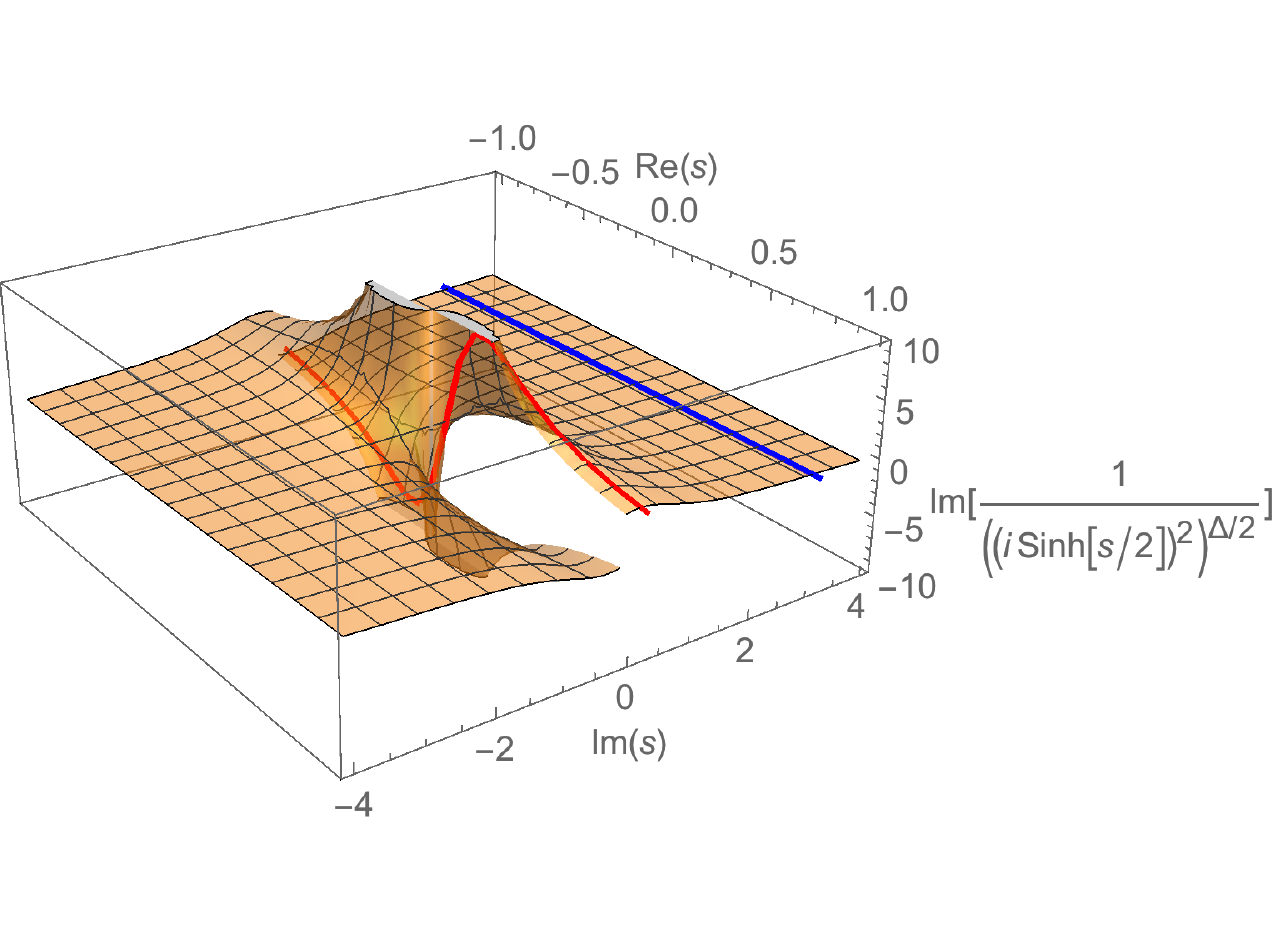}
\caption{Branch cut structure of the correlator \eqref{eq:3ptfunc}, along with the original (red) and deformed (blue) contour in \eqref{eq:triop1a}.}
\label{fig:1}
\end{figure}

Here we would like to reproduce the coefficient of trioperator exchange, i.e. $\theta_0^{3\Delta}$ in the small subsystem limit, $B_O( \tau,\hat\tau) \approx (2\theta_0)^{\Delta} O(\hat\tau) + \cdots$.
In this case, we need the three point function on $\Sigma_1=S^1 \times H^{d-1}$. This is easy to calculate using the map  \eqref{eq:hyptoflat}. The result is\footnote{
We have collapsed powers $\sqrt{\vec{x}^2}$ for terms like the ones involving $\cosh$, which are positive along the integration contours. We have kept powers explicit for the $\sinh$ to make manifest the branch cut structure in the $v=s_2-s_1$ plane. There is a cut from $v=-\infty$ to $v=0$ and one from $v=0$ to $v=\infty$ and they join in the light cone singularity at $v=0$. According to sec. \ref{sec:contourp3}, the contour must run on the upper half plane, i.e. it does not cross the branch cut anywhere, see Fig. \ref{fig:1} for an illustration.} 
\beq
\label{eq:3ptfunc}
\langle   O(\hat \tau-\tau_{s_1}) O(\hat \tau-\tau_{s_2})  O(\hat \tau)  \rangle_{\Sigma_1} = \frac{C_{OOO}}{\Big[8\cosh \frac{s_1}{2}\cosh \frac{s_2}{2}\sqrt{(i\sinh \frac{s_2-s_1}{2})^2}\Big]^\Delta} + O(\theta_0),
\eeq
so that by putting this into \eqref{eq:cubicee} we get
\bea
\label{eq:triop0}
\delta S_V^{(3)} &= -(2\theta_0)^{3\Delta} (C^{O}_{VV})^3\frac{ C_{OOO} }{16\pi 8^\Delta }\int_{-\infty}^{\infty} ds_1ds_2 \frac{i(s_2-s_1)}{2\pi}\\
&\times \left( \frac{i}{(\cosh \frac{s_1}{2}\cosh \frac{s_2}{2})^{\Delta+1}\sinh \frac{s_2-s_1}{2} \left[ (i\sinh \frac{s_2-s_1}{2})^2\right]^{\Delta/2}} \right).
\eea
The integral is easy to evaluate numerically provided that we shift $s_i \rightarrow s_i-i\epsilon_i$, with $\pi>\epsilon_1>\epsilon_2>-\pi$ (see discussion in section \ref{sec:contourp3}). For well-separated $\epsilon_i$s, the integral has good convergence properties and the result does not depend on the choice of $\epsilon_i$s. Here we carry on to derive the analytic value of the integral. This is slightly convoluted, so we advise the reader not interested in this to jump directly to formula \eqref{eq:triopfinal}.

To evaluate \eqref{eq:triop0}, we first need to evaluate the auxiliary integral
\beq
Q=(2\theta_0)^{3\Delta} (C^{O}_{VV})^3\frac{ C_{OOO} }{16\pi 8^\Delta }\int_{-\infty}^{\infty} ds_1ds_2\left( \frac{i}{(\cosh \frac{s_1}{2}\cosh \frac{s_2}{2})^{\Delta+1}\sinh \frac{s_2-s_1}{2} \left[ (i\sinh \frac{s_2-s_1}{2})^2\right]^{\Delta/2}} \right),
\eeq
which differs from $\delta S_V^{(3)}$ by not having the linear terms.
First, let us assume $\Delta$ to be an even positive integer, in which case the branch cuts in the integrand disappear. In this case, the powers can be collapsed and we can evaluate the integral using (\ref{eq:auxint1}-\ref{eq:int2b})
\bea
\label{eq:triop1}
Q &=(2\theta_0)^{3\Delta} (C^{O}_{VV})^3\frac{ C_{OOO} }{16\pi 8^\Delta } \int_{-\infty}^{\infty} ds_1ds_2 \left( \frac{i}{\cosh \frac{s_1}{2}\cosh \frac{s_2}{2}\sinh \frac{s_2-s_1}{2}} \right)^{\Delta+1} \\ &=(2\theta_0)^{3\Delta} (C^{O}_{VV})^3C_{OOO}\frac{\Gamma(\frac{1+\Delta}{2})^3}{4\pi \Gamma(\frac{3+3\Delta}{2})}.
\eea
For general $\Delta$, we first decouple directly the $s_1$ and $s_2$ integrals by inserting $\int du \delta(u-(s_2-s_1))$ and then pass over to Fourier representation of the delta
\bea
\label{eq:triop1a}
\int_{-\infty}^{\infty} ds_1ds_2 &\left( \frac{i}{(\cosh \frac{s_1}{2}\cosh \frac{s_2}{2})^{\Delta+1}\sinh \frac{s_2-s_1}{2} \left[ (i\sinh \frac{s_2-s_1}{2})^2\right]^{\Delta/2}} \right)  \\
&=
i\int \frac{dq}{2\pi} \left[ \int_{-\infty}^\infty ds \frac{e^{iq s}}{(\cosh\frac{s}{2})^{\Delta+1}}\right]^2 \left[ \int du \frac{ e^{iq u}}{\sinh\frac{u}{2}((i\sinh\frac{u}{2})^2)^{\Delta/2}}\right]\\
&=\int \frac{dq}{2\pi}\left[ \int_{-\infty}^\infty ds \frac{e^{iq s}}{(\cosh\frac{s}{2})^{\Delta+1}}\right]^2 \left[ \int du \frac{  e^{-\pi q}e^{iq u}}{(\cosh\frac{u}{2})^{\Delta+1}}\right],
\eea
In the last line, we have deformed the contour as $u\rightarrow u+i\pi$, remembering that the $u$ integral had to run above the branch cuts and the singularity, see Fig. \ref{fig:1} and the discussion in section \ref{sec:contourp3}. This way we can now collapse the powers as $\cosh\frac{u}{2}$ is positive along the new contour. We obtain the formula
\beq
\label{eq:triop1b}
Q=(2\theta_0)^{3\Delta} (C^{O}_{VV})^3\frac{ C_{OOO} }{16\pi 8^\Delta }\int \frac{dq}{2\pi}e^{-\pi q}\left[ \int_{-\infty}^\infty ds \frac{e^{iq s}}{(\cosh\frac{s}{2})^{\Delta+1}}\right]^3, 
\eeq
 The $s$ integral is easy to evaluate by a change of variable
\beq
\int_{-\infty}^\infty ds \frac{e^{iq s}}{(\cosh\frac{s}{2})^{\Delta+1}} = \frac{2^{1+\Delta}\Gamma(\frac{1+\Delta}{2}+iq)\Gamma(\frac{1+\Delta}{2}-iq)}{\Gamma(\Delta+1)}.
\eeq
We do not know how to do the resulting $q$ integral analytically, but it is now an integral with very good convergence properties, and can easily be evaluated numerically for any $\Delta>0$. Doing so, we get results for \eqref{eq:triop1b} which are equal to \eqref{eq:triop1} (peviously valid for $\Delta \in 2 \mathbb{Z}$) for any $\Delta$.

Now we move on to compute the actual $\delta S^{(3)}_V$ in \eqref{eq:triop0} and relate it to $Q$ in \eqref{eq:triop1}.
Let us do the same trick of decoupling the integrals by the insertion of the delta function:
\bea
\label{eq:triop2}
\int_{-\infty}^{\infty} ds_1ds_2 \frac{i(s_2-s_1)}{2\pi}&\left( \frac{i}{(\cosh \frac{s_1}{2}\cosh \frac{s_2}{2})^{\Delta+1}\sinh \frac{s_2-s_1}{2} \left[ (i\sinh \frac{s_2-s_1}{2})^2\right]^{\Delta/2}} \right)\\
&=
i\int \frac{dq}{2\pi} \left[ \int_{-\infty}^\infty ds \frac{e^{iq s}}{(\cosh\frac{s}{2})^{\Delta+1}}\right]^2 \left[ \int du \frac{ \frac{i u}{2\pi}e^{iq u}}{\sinh\frac{u}{2}((i\sinh\frac{u}{2})^2)^{\Delta/2}}\right]\\
&=i\int \frac{dq}{2\pi} \left[ \int_{-\infty}^\infty ds \frac{e^{iq s}}{(\cosh\frac{s}{2})^{\Delta+1}}\right]^2 \frac{\partial_q}{2\pi}\left[ \int du \frac{ e^{iq u}}{\sinh\frac{u}{2}((i\sinh\frac{u}{2})^2)^{\Delta/2}}\right]\\
&=\int \frac{dq}{2\pi}\left[ \int_{-\infty}^\infty ds \frac{e^{iq s}}{(\cosh\frac{s}{2})^{\Delta+1}}\right]^2 \frac{\partial_q}{2\pi}\left[ \int du \frac{  e^{-\pi q}e^{iq u}}{(\cosh\frac{u}{2})^{\Delta+1}}\right],
\eea
where in the third line we have replaced the $\frac{i u}{2\pi}$ factor by a $\frac{1}{2\pi}\partial_q$ and in the last line we have again deformed the $u$ contour as $u\rightarrow u+i\pi$.
Had we not have the linear factor, we would have had an identical expression as \eqref{eq:triop1b}. 
On the other hand, we can rewrite the last line of \eqref{eq:triop2} as
\beq
\frac{1}{2\pi}\int \frac{dq}{2\pi} e^{2\pi q} \frac{1}{3}\partial_q\left[ e^{-\pi q}\int_{-\infty}^\infty ds \frac{e^{iq s}}{(\cosh\frac{s}{2})^{\Delta+1}}\right]^3,
\eeq
which, after partial integration, gives
\beq
-\frac{1}{3}\int \frac{dq}{2\pi} e^{-\pi q} \left[\int_{-\infty}^\infty ds \frac{e^{iq s}}{(\cosh\frac{s}{2})^{\Delta+1}}\right]^3,
\eeq
which is just $-1/3$ times the integral in \eqref{eq:triop1b}. Therefore, the final result in the small subsystem limit is just $1/3$ times the auxiliary integral of \eqref{eq:triop1}, i.e.
\beq
\label{eq:triopfinal}
\delta S_V^{(3)} =-\left( -\frac{1}{3} Q \right)= (2\theta_0)^{3\Delta} (C^{O}_{VV})^3 C_{OOO} \frac{\Gamma(\frac{1+\Delta}{2})^3}{12\pi \Gamma(\frac{3+3\Delta}{2})}.
\eeq
This agrees with the result of the holographic computation \cite{Casini:2016rwj,Sarosi:2016atx}\footnote{There is an unfortunate mistake in \cite{Sarosi:2016atx}, which leads to a discrepancy with e.q. (5.5) there. The mistake comes from a typo in (5.4), which should correctly be
\beq
\nonumber
\kappa = -\frac{1}{\eta} \frac{2\pi^d \Gamma(\Delta-\frac{d}{2})^3}{\Gamma(\frac{\Delta}{2})^3\Gamma(\frac{3\Delta-d}{2})} C_{\tilde O \tilde O \tilde O},
\eeq
Taking this into account, the result matches \eqref{eq:triopfinal}.
} and provides a direct example where the derivation was so far not accessible in the CFT using the replica trick.\footnote{However, for $\Delta=2$ see \cite{He:2017vyf}.}


\section{Discussion and outlook}
\label{sec:discuss}

In this work, we have developed a new technique to calculate the modular Hamiltonian and the entanglement entropy for slightly excited states reduced to a ball shaped region in conformal field theories. The method is based on a suggestive rewriting of the Taylor expansion of the noncommutative logarithm $\log(\rho_0+\delta \rho)$ in terms of powers of $\delta \rho$, translated and integrated along the modular flow of the reference state $\rho_0$. For the entanglement entropy (or also e.g. for the relative entropy) in CFT, this results in an expansion in which the $n$th term is given by a $2n+2$ point function of the operator creating the excited state, integrated along the vacuum modular flow. The expansion acquires a small control parameter in different situations, e.g. this can be the size of the subsystem, or $1/N$ in holographic theories (modulo the contribution of multitrace operators, which we soon discuss), or a probability parametrizing the mixture of the vacuum and the excited state reduced density matrices. We have demonstrated the usefulness of the method by reproducing some earlier results in the small subsystem size expansion -- some of which only had a derivation from holography so far -- with relative ease. 

In addition, we have provided a direct holographic rewriting of the first ($n=1$) term in this expansion in terms of an emergent bulk field defined via the HKLL construction. The first term is just the canonical energy of this free bulk field, calculated on the codimension 1 surface between the boundary region and the RT surface in AdS. The novel feature of this statement is that it holds for any CFT without assuming large $N$ or the Ryu-Takayanagi formula.

Below we summarize some open questions and possible future directions. 

\subsubsection*{Resumming the entropy}

It would be interesting to find instances when the formula (\ref{eq:entropyexpansion1},\ref{eq:entropyexpansion}) can be resummed, as this would give an exact expression. One possible instance is the following.
As we have mentioned previously, when $V$ is a light single trace operator, the OPE coefficients appearing in the expansion of the entanglement entropy are $\frac{1}{N}$ suppressed, apart from the multitrace contributions from $V$, like fusion into $\om = : VV: $. These come with the three point function coefficient $C_{VV \om}$ which is $o(N^0)$. It is easy to see that these OPE coefficients are responsible for giving the generalized free field result in multipoint correlators. Resumming this contribution therefore would first require one to evaluate the integrals in \eqref{eq:entropyexpansion}, with the multipoint functions calculated via all Wick-contractions, as in free field theory. It is an interesting question for the future whether this can be done.\footnote{We thank Alexandre Belin, Nabil Iqbal and Sagar Lokhande for discussions on this, and for sharing some preliminary results.} We note that in two dimensions, the result is known for operators with conformal weights $(h,\bar h)=(1,0)$, as it can be calculated using the replica trick \cite{2014JSMTE..09..025C,2013PhRvL.110k5701E,Ruggiero:2016khg}. 

\subsubsection*{Holographic interpretation for higher order terms}

Given the holographic interpretation of the leading terms in the expansion, it would be interesting to see how, and under what assumptions, other terms can be interpreted holographically. It should be possible to explicitly find the effect of bulk interactions, gravitational dressing,\footnote{In particular, it would be interesting to study the replacement of global OPE blocks with Virasoro blocks in two dimensions, along the lines of \cite{Guica:2016pid}.} and loop corrections and thus compare to the predictions of \cite{Faulkner:2013ana}. 

\subsubsection*{General backgrounds}

One might be interested in small perturbations of a state which is not the vacuum. In holography, we could have a heavy state, creating a classical background geometry, and excitations of some bulk quantum fields on top of it in mind. The Rindler representation \eqref{eq:rhoV} is still valid if $|V\rangle$ is a heavy state. Let us assume $|V'\rangle$ is some small excitation of $|V\rangle$ and write
\bea
\rho_{V'} &= \frac{ e^{-\pi K} V'( \tau-\pi)  V'(\hat \tau-\pi)e^{-\pi K} }{\langle V'(\tau)V'(\hat \tau)\rangle_{S^1\times H^{d-1}}} \\
&=\rho_V + e^{-\pi K}\sum_{k\neq vac} (C^k_{V'V'}-C^k_{VV})B_k e^{-\pi K},
\eea
so that we can use
\beq
\delta \rho = e^{-\pi K}\sum_{k\neq vac} (C^k_{V'V'}-C^k_{VV})B_k e^{-\pi K},
\eeq
in \eqref{eq:generalform}, but we need to replace the vacuum modular flow by the modular flow of $\rho_V$. This gives a formal expansion in $(C^k_{V'V'}-C^k_{VV})$, which is the background substracted one point function. Now most of the considerations of this paper cannot be directly applied for this expansion. The reason is that the modular flow of $\rho_V$ is non-geometric, and the conjugated operators in \eqref{eq:generalform} become very complicated. However, it was recently pointed out in \cite{Jafferis:2015del,Faulkner:2017vdd} that for holographic theories where $|V\rangle$ creates a classical background, the modular flow with $\rho_V$ on the boundary is the same as the vacuum modular flow in the bulk, in the backreacted geometry (at least to leading order in $1/N$). This was recognized to be a key feature for the emergence of bulk locality in \cite{Faulkner:2017vdd}. This way for holographic theories, some terms of our expansion can be written again in a simple form in terms of bulk fields 
and a formula similar to \eqref{eq:entropyexpansion} could be written down. We leave these matters to possible future work.

\section*{Acknowledgements}
We are grateful to Cesar Ag\'on, Alexandre Belin,  Simon Gentle, Nabil Iqbal, Aitor Lewkowycz, Sagar Lokhande, Javier Mag\'an, Tadashi Takayanagi and Eduardo Test\'e for discussions. We thank Alexandre Belin and Onkar Parrikar for comments on the manuscript. We also thank the Yukawa Institute (YITP) for hospitality, where a part of this work was done. 
The work of G.S. was supported in part by a grant from the Simons Foundation (\#385592, Vijay Balasubramanian) through the It From Qubit Simons Collaboration, by the Belgian Federal Science Policy Office through the Interuniversity Attraction Pole P7/37, by FWO-Vlaanderen through projects G020714N and G044016N, and by Vrije Universiteit Brussel through the Strategic Research Program ``High-Energy Physics''. The work of T. U. was supported in part by the National Science Foundation under Grant
No. NSF PHY-25915. 

\appendix

\section{Coordinates of the hyperbolic black hole} 
\label{section:hb}

Following \cite{Faulkner:2014jva} here we summarize the properties of  the hyperbolic black hole mainly to fix the notation.  We start from $d+1$ dimensional Anti de Sitter space in the $d+2$ dimensional  embedding space 
\be
-(X_{0})^2 +(X_{d+2})^{2} +(X_{i})^2=-1, \quad i=1 \cdots d. 
\ee
The Euclidean hyperbolic black hole metric of AdS is given by 
\be
ds^2= (r^2-1) d\tau^2 +\f{dr^2}{r^2-1} +r^2  (-dY_{I}^2+ dY_{a}^2),
\ee 
 where $(Y_{I}, Y_{a}),  $ are the coordinates of the $d-1$ dimensional hyperbolic space $H^{d-1}$ with the constraint $ -(Y_{I})^2+ (Y_{a})^2=-1$. 
This  metric is coming from the embedding   
\be
\left(X_{0}, X_{d+2}, X_{i} \right) = \left( rY^{I}, \s{r^2-1}\cos \tau, \s{r^2-1}\cos \tau,  r Y_{a} \right).
\ee
The boundary of the hyperbolic black hole is $S^{1} \times H^{d-1}$. In the embedding space this is realized as the null projective cone
\be
P^2=0, \quad P =\lim_{r \rightarrow \infty } X/r .
\ee
In the body of the paper we also need the coordinates $X_{B}:(l_{B}, Y_{B})$ of the future horizon of the black hole, which is derived by first going to Lorentzian regime $\tau \rightarrow is $ and then taking the near horizon limit $ s \rightarrow \infty, r\rightarrow 1$ while keeping $l_{B} =e^{s} \s{r^2-1}$ fixed 
\be
X_{B} = \left(Y^{I}_{B}, \f{l_{B}}{2}, -i\f{l_{B}}{2}, Y^{m}_{B} \right).
\ee
In the embedding space description, the bulk to boundary propagator  $\la \phi (X) \om (P) \ra$ of  the free scalar field $\phi (X)$  is simply 
\be
\la \phi (X) \om (P) \ra = \f{c_{\Delta}}{(-2X \cdot P)^{2\Delta}}, \quad c_{\Delta} =\f{2(\Delta-\f{d}{2} )\Gamma(\Delta)}{\pi^{\f{d}{2}}\Gamma(\Delta-\f{d}{2})},
\ee
 where $\Delta$ is the conformal dimension of the boundary operator $\mathcal{O}$ which is related to the mass $m$ of the bulk field by $m^2= \Delta (\Delta-d)$.
In particular, when the bulk point is on the future horizon $(l_{B}, Y_{B})$, and the boundary point $(\tau_{a}, Y_{a})$ is on the Euclidean section $ S^{1} \times H^{d-1}$,  the propagator is 
\be 
\la \phi (l_{B},Y_{B}) \om (\tau_{a}, Y_{a}) \ra= \f{c_{\Delta}}{\left(-2Y_{B}\cdot Y_{a} -l_{B} e^{-i \tau_{a}} \right)^{\Delta}}.
\ee

\section{Some useful integrals}
\label{app:1}

We will use the expression
\beq
\label{eq:int1}
\int_{-\infty}^\infty ds \frac{1}{(2\cosh\frac{s}{2})^\nu} = \frac{\sqrt{\pi}\Gamma(\frac{\nu}{2})}{2^{\nu-1}\Gamma(\frac{1}{2}\nu+\frac{1}{2})}.
\eeq
We will also need
\beq
\label{eq:auxint1}
I(\nu,\mu)=\int_{-\infty}^{\infty} ds_1ds_2 (\cosh \frac{s_1}{2}\cosh \frac{s_2}{2})^\mu\left( \frac{1}{\cosh \frac{s_1}{2}\cosh \frac{s_2}{2}\sinh \frac{s_2-s_1}{2}} \right)^{\nu+1}
\eeq
with the prescription that we go around the singularity $s_1=s_2$ as $s_2-s_1 \rightarrow s_2-s_1+i\epsilon$, $\epsilon>0$. Assume $\nu \in \mathbb{Z}$ and $\nu>-1$, $\mu-\nu<1$. First we exchange integration variables to $s_i=2 \text{arctanh}(2v_i)$, $i=1,2$:
\beq
I(\nu,\mu)=(-1)^{-\mu}2^{3-\nu} \int_{-1/2}^{1/2}dv_1\int_{-1/2}^{1/2}dv_2\frac{\left((4v_1^2-1)(4v_2^2-1) \right)^{\nu-\frac{\mu}{2}}}{(v_2-v_1)^{\nu+1}}
\eeq
We further write
\bea
I(\nu,\mu) &= (-1)^{-\mu}2^{3-\nu}\int_{-\infty}^\infty du\int_{-1/2}^{1/2}dv_1\int_{-1/2}^{1/2}dv_2\frac{\left((4v_1^2-1)(4v_2^2-1) \right)^{\nu-\frac{\mu}{2}}}{u^{\nu+1}}\delta(u-v_2+v_1)\\
&=(-1)^{-\mu}2^{3-\nu}\int_{-\infty}^\infty \frac{dq}{2\pi} \left[\int_{-1/2}^{1/2}dv(4v^2-1)^{\nu-\frac{\mu}{2}} e^{-iq v}\right]^2 \left[\int_{-\infty}^\infty du\frac{1}{u^{\nu+1}}e^{iq u}\right],
\eea
where we have decoupled the integrals inside the $q$ integral. Assuming $\nu \in \mathbb{Z}$, we can do the $u$ integral by closing the contour (and recalling that our prescription requires going above the singularity):
\beq
\int_{-\infty}^\infty du\frac{1}{u^{\nu+1}}e^{iq u} = -\frac{2\pi i^{\nu+1}q^\nu}{\Gamma(\nu+1)}\Theta(-q),
\eeq
where $\Theta$ is the usual Heaviside theta function. 
We can directly evaluate the other integral given $\nu-\frac{\mu}{2}>-1$:
\beq
\int_{-1/2}^{1/2}dv(4v^2-1)^{\nu-\frac{\mu}{2}} e^{-iq v} = (-1)^{\nu-\frac{\mu}{2}}\sqrt{\pi}\Gamma(\nu-\frac{\mu}{2}+1) 2^{2(\nu-\frac{\mu}{2})}q^{-\frac{1}{2}-\nu+\frac{\mu}{2}} J_{\frac{1}{2}+\nu-\frac{\mu}{2}}(q/2).
\eeq
Here, $J_n(x)$ is Bessel's function of the first kind. This way we arrive at the integral
\bea
\label{eq:int2b}
I(\nu,\mu) &= (-1)^{1-\mu}2^{4-2\mu+3\nu}i^{\nu+1}\pi^2 \int_{-\infty}^0 \frac{dq}{2\pi} \left[ J_{\frac{1}{2}+\nu-\frac{\mu}{2}}(\frac{q}{2}) \right]^2 q^{\mu-\nu-1}\\
&=(-1)^{1-\mu-\nu}\frac{i^{1+5\nu}2^{2-\mu+\nu} \pi \Gamma(1-\frac{\mu}{2}+\nu)^2 \Gamma(\frac{1-\mu+\nu}{2})}{\Gamma(\frac{2-\mu+\nu}{2})\Gamma(\frac{3}{2}-\mu+\frac{3\nu}{2})\Gamma(\frac{\nu}{2}+1)}.
\eea
where we have again assumed $\nu$ to be integer. 

\section{On the divergence at $\theta_0=\pi/2$}
\label{app:3}

Let us examine Fig. \ref{fig:2}, representing the Euclidean projection of the operator insertions, for the case $n=1$. There is a single $\epsilon_1$ and as $\theta_0$ grows, the best we can choose to keep the operator ordering is $\epsilon_1=0$. As $\theta_0\rightarrow \pi/2$, the operator inserted at $\tau=\pi-\theta_0$ collides with the one at $\tau=\theta_0+i s_1$ at the point $s_1=0$ along the integral, and there is no two point function factor in \eqref{eq:entropyexpansion} canceling the divergence coming from this. There is of course a similar problem with the pair at $-\pi+\theta_0$ and $-\theta_0+i s_1$. The integral contour therefore approaches a power law singularity, which in general leads to a divergence as $\theta_0 \rightarrow \pi/2$. To show that this really happens, we now examine this integral for chiral vertex operators in the $d=2$ free scalar CFT.

We want to evaluate
\bea
\langle V|\delta K^{(1)}|V\rangle &=\int ds \mathcal{K}_1(s) \left\langle \frac{V_{\alpha}(\tau)V_{-\alpha}(\hat \tau)}{\langle V_{\alpha}(\tau)V_{-\alpha}(\hat \tau) \rangle} \left( \frac{V_\alpha(\tau-\tau_s)V_{-\alpha}(\hat \tau-\tau_s)}{\langle V_\alpha(\tau)V_{-\alpha}(\hat \tau) \rangle}-1\right) \right\rangle \\
&= \int ds \mathcal{K}_1(s)\left\langle \frac{V_\alpha(\tau)V_{-\alpha}(\hat \tau)}{\langle V_\alpha(\tau)V_{-\alpha}(\hat \tau) \rangle} \frac{V_{\alpha}(\tau-\tau_s)V_{-\alpha}(\hat \tau-\tau_s)}{\langle V_{\alpha}(\tau)V_{-\alpha}(\hat \tau) \rangle}\right\rangle -1.
\eea
We have used $\int ds \mathcal{K}_1(s)=1$ here. For vertex operators on the cylinder we have
\bea
\langle  V_\alpha(\tau)V_{-\alpha}(\hat \tau) V_{\alpha}(\tau-\tau_s)V_{-\alpha}(\hat \tau-\tau_s) \rangle  
&= \left[ 2\sin(\frac{\tau-\hat \tau}{2})\right]^{-2\alpha^2} \left[ 2\sin(\frac{\tau_s}{2})\right]^{2\alpha^2}\\ &\times \left[ 2\sin(\frac{\tau-\hat \tau+\tau_s}{2})\right]^{-\alpha^2}\left[ 2\sin(\frac{\hat \tau- \tau+\tau_s}{2})\right]^{-\alpha^2} \\
\langle V_\alpha(\tau)V_{-\alpha}(\hat \tau) \rangle &= \left[ 2\sin(\frac{\tau-\hat \tau}{2})\right]^{-\alpha^2}.
\eea
Therefore, after simplification
\beq
\left\langle \frac{V_\alpha(\tau)V_{-\alpha}(\hat \tau)}{\langle V_\alpha(\tau)V_{-\alpha}(\hat \tau) \rangle} \frac{V_{\alpha}(\tau-\tau_s)V_{-\alpha}(\hat \tau-\tau_s)}{\langle V_{\alpha}(\tau)V_{-\alpha}(\hat \tau) \rangle}\right\rangle = \left[ 2\cosh(\frac{s}{2})\right]^{2\alpha^2}\left[ 2(2\cosh^2 \frac{s}{2}-2\sin^2\theta_0)\right]^{-\alpha^2}.
\eeq
We will follow the safest approach and expand in $q=-4\sin^2\theta_0$. This roughly amounts to doing the integrals with an unsummed OPE, and then resumming the result. We can use the generalized binomial expansion to write
\bea
\left\langle \frac{V_\alpha(\tau)V_{-\alpha}(\hat \tau)}{\langle V_\alpha(\tau)V_{-\alpha}(\hat \tau) \rangle}  \frac{V_{\alpha}(\tau-\tau_s)V_{-\alpha}(\hat \tau-\tau_s)}{\langle V_{\alpha}(\tau)V_{-\alpha}(\hat \tau) \rangle}\right\rangle &= \left[ 2\cosh(\frac{s}{2})\right]^{2\alpha^2} \\ &\times \sum_{k=0}^\infty \binom{-\alpha^2}{k}\left[ 2\cosh(\frac{s}{2})\right]^{-2\alpha^2-2 k} (-4 \sin^2 \theta_0)^{k},
\eea
where
\beq
\binom{-\alpha^2}{k} = \frac{\Gamma(1-\alpha^2)}{\Gamma(k+1)\Gamma(1-k-\alpha^2)}.
\eeq
The appearing integral is of the same form as in \eqref{eq:int1}.
Therefore,
\bea
\label{eq:vertex}
\langle V|\delta K^{(1)}|V\rangle &= -1+\sum_{k=0}^\infty \frac{2^{-1-2k} \sqrt{\pi}\Gamma(1-\alpha^2)}{\Gamma(\frac{3}{2}+k)\Gamma(1-\alpha^2-k)}(-4 \sin^2 \theta_0)^{k} \\
&= {}_2 F_1 \left( 1,\alpha^2;\frac{3}{2}; \sin^2 \theta_0 \right) -1.
\eea
This is singular as $\theta_0 \rightarrow \pi/2$ whenever $\alpha^2 \geq 1/2$.

In principle, there could be similar singularities appearing in the $n$th term \eqref{eq:entropyexpansion} for $\theta_0=\pi/(k+1)$, $k=1,...,n$. Now both colliding operators have Lorentzian shifts, so one meets integrals of the form
\beq
\int ds_k ds_{k'} \frac{1}{(s_k-s_{k'})^\mu}.
\eeq
The double integral improves convergence properties, but in general can still lead to divergences. We did not manage to perform the $n=2$ integrals for vertex operators to provide an explicit example. We leave this as an open question.

\bibliographystyle{utphys}
\bibliography{opeblock}

\end{document}